\documentclass[sigconf,nonacm]{acmart}

\AtBeginDocument{%
  }

\setcopyright{acmlicensed}
\copyrightyear{2026}
\acmYear{2026}
\acmDOI{XXXXXXX.XXXXXXX}

\acmConference[ACM DIS '26]{ACM Designing Interactive Systems}{June 13--17,
  2026}{Singapore}

\acmISBN{978-1-4503-XXXX-X/2018/06}

\usepackage{multirow}
\usepackage{rotating}
\usepackage{tabularx}
\usepackage{adjustbox}

\newcommand{\toollike}{~\emph{tool-like~}}

\begin{document}

\title[From Retrieving Information to Reasoning with AI]{From Retrieving Information to Reasoning with AI: Exploring Different Interaction Modalities to Support Human-AI Coordination in Clinical Decision-Making}

\author{Behnam Rahdari}
\affiliation{
  \institution{Stanford University}
    \city{Palo Alto}
  \state{CA}
  \country{USA}  
}
\email{behnamr@stanford.edu}

\author{Sameer Shaikh}
\affiliation{
  \institution{Joseph Brant Hospital \& Foundation}
    \city{Ontario}
  \state{ON}
  \country{Canada}  
}
\email{sameer@inflectiveai.co}

\author{Jonathan H Chen}
\affiliation{
  \institution{Stanford University}
    \city{Palo Alto}
  \state{CA}
  \country{USA}  
}
\email{jonc101@stanford.edu}

\author{Tobias Gerstenberg}
\affiliation{
  \institution{Stanford University}
    \city{Palo Alto}
  \state{CA}
  \country{USA}  
}
\email{gerstenberg@stanford.edu}

\author{Shriti Raj}
\affiliation{
  \institution{Stanford University}
    \city{Palo Alto}
  \state{CA}
  \country{USA} 
}
\email{shritir@stanford.edu}

\renewcommand{\shortauthors}{Rahdari et al.}

\begin{abstract}
  LLMs are popular among clinicians for decision-support because of simple text-based interaction. However, their impact on clinicians’ performance is ambiguous. Not knowing how clinicians use this new technology and how they compare it to traditional clinical decision-support systems (CDSS) restricts designing novel mechanisms that overcome existing tool limitations and enhance performance and experience. This qualitative study examines how clinicians (n=12) perceive different interaction modalities (text-based conversation with LLMs, interactive and static UI, and voice) for decision-support. In open-ended use of LLM-based tools, our participants took a tool-centric approach using them for information retrieval and confirmation with simple prompts instead of use as active deliberation partners that can handle complex questions. Critical engagement emerged with changes to the interaction setup. Engagement also differed with individual cognitive styles. Lastly, benefits and drawbacks of interaction with text, voice and traditional UIs for clinical decision-support show the lack of a one-size-fits-all interaction modality.

\end{abstract}



\maketitle

\section{Introduction}

Recent advances in large language models (LLMs) have extended artificial intelligence in medicine beyond narrow pattern recognition toward multi step clinical reasoning. Across multiple domains, these advanced AI systems now reach diagnostic performance comparable to or exceeding that of less experienced clinicians \cite{shen2019artificial}. Large language model based systems further demonstrate specialist--level differential diagnostic reasoning and can improve clinician performance on challenging real world cases \cite{mcduff2025}. These capabilities are beginning to move AI from a back end analytic tool into the flow of clinical work itself \cite{topol2019high}.

Recent evidence indicates that large language models can improve physician performance on management and treatment planning tasks. In a randomized controlled trial with 92 practicing physicians, GPT-4 was evaluated on open ended clinical management reasoning, including balancing diagnostic tests, treatments, and risk under uncertainty \cite{goh2025gpt}. Physicians using GPT-4 alongside conventional resources achieved higher expert-scored performance than those using conventional resources alone, while also spending more time per case. At the same time, GPT-4 operating without a physician performed at a level statistically indistinguishable from GPT-4-assisted clinicians, showing that the model itself was capable of carrying out the management reasoning task at a high level.

What remains unclear is how clinicians and models coordinate during decision making and when and how these models add value for clinicians. Research on human–AI teaming shows that performance depends not only on the capabilities of each party, but on how roles, communication, and shared understanding are structured in the interaction \cite{andrews2023role, Berretta2023}. Human–AI collaboration research further emphasizes that effective teaming requires explicit support for coordination and mutual interpretation rather than treating the AI as a standalone tool \cite{Lou2025}. At the same time, existing evaluation methods for dialogue systems largely focus on task success or response quality, and provide limited insight into how joint reasoning and alignment are produced during interaction \cite{Deriu2021}. As a result, although we now know that LLMs can influence clinical performance, we still lack empirical understanding of how collaboration with these systems is actually enacted in practice and where such collaboration fails.

Clinical decision making is not only a matter of producing the correct answer, but of doing so under time pressure, uncertainty, and cognitive strain. Decades of work show that clinicians operate in environments characterized by information overload, competing goals, and the need to commit before all evidence is available, which creates systematic vulnerability to bias and premature closure \cite{PhillipsWren2020}. In these settings, errors arise not from lack of knowledge but from how attention and judgment are managed in the moment \cite{Croskerry2003}. The introduction of AI systems into this context changes the structure of cognitive work, but not the underlying constraints and goals. Studies of AI-assisted decision making show that users often only briefly attend to AI outputs, and adopt heuristics about when to follow or ignore model outputs rather than carefully evaluating each recommendation, which can lead to overreliance or automation bias even when explanations are provided \cite{Bucinca2021, Lyell2017}. This makes the design of interaction with AI a critical facilitator of clinical safety and performance, not just the quality of the model’s predictions.

Most human–AI interfaces for large language models are built on a conversational chat metaphor, inherited from earlier work on messaging bots and conversational agents \cite{Klopfenstein2017, Diederich2022}. This is a significant departure from pre-LLM forms of clinical decision-support that primarily involved machine-learned recommendations presented through a user interface or alerts. While traditional AI/ML based decision-support tools did not see much success outside the lab, the popularity of LLMs among clinicians warrants understanding how their use differs from other types of decision-support interfaces that clinicians interact with so that appropriate human-AI interactive experiences can be designed .  

This paper examines how clinicians (n=12) experience AI support in working through complex clinical vignettes across three forms of interaction: free text conversation with a large language model, a UI representing AI decision-making and recommendation, and voice based systems, such as ambient scribes used in practice. The study focuses on how interaction formats shape what AI-generated information clinicians read, trust, discard, and act on during clinical decision making. 
We found that conversational interaction supported rapid information lookup but rarely triggered or supported collaborative reasoning and deliberation between clinicians and AI. Positioning the LLM in a specific role, such as a specialist, and scaffolding prompt creation by anchoring in it familiar text-based activity resulted in deeper engagement with AI. Visual UI artifacts externalized reasoning in a form that enabled relatively more critical engagement through quick inspection, comparison, and critique of AI's reasoning. Voice based interaction was perceived as disruptive for attention management and was considered less practical for creating and consuming decision-support artifacts. Across modalities, we observed a dislike for probabilistic language and layered or nested interfaces. Lastly, visual representations of reasoning that matched clinicians' own cognition were preferred over text-based representations. Drawing on our findings, we offer implications for designing clinical decision-support tools to support clinicians' attention management across different information sources and for multitasking, add interactivity without changing the information space through all-in-one interfaces, and support coordinated use of different modalities (e.g., voice for input, UIs for reasoning).

These findings indicate that whether an LLM becomes a source of facts or a partner in reasoning is not determined by model capability alone, but by how interaction structures shape what parts of a decision are made visible, comparable, and open to challenge. In our study, clinicians readily provided rich case context through conversation, yet still treated the model as a lookup tool by selectively scanning its output, extracting a single confirming cue, and moving on. Deeper engagement emerged only when interaction formats changed how AI outputs were represented and worked with, such as when reasoning was externalized into stable visual artifacts or when the model was positioned as a specialist needing a formal clinical note as an input. Our findings suggest that in high-stakes clinical work, where attention is limited and accountability is high, supporting collaborative decision making will require representing AI reasoning in formats that support assessment of multiple factors at once and are aligned with how clinicians think. Additionally, such representations should enable non-linear information consumption and interaction that allows clinicians to split attention in naturalistic ways.

Our work contributes an empirical understanding of how clinicians engage with LLMs to solve complex clinical vignettes and tasks and how this experience compares with the use of UI and voice modalities. It also contributes design implications for clinical decision-support tools.

\section{Related Work}
This work sits at the intersection of human--AI teaming, large language models (LLMs) in clinical decision support, and clinician-facing interaction design. Prior research offers (i) conceptual models and metrics for human--AI teams, (ii) growing evidence about LLM capabilities to support clinicians and their deployment constraints in healthcare, and (iii) HCI approaches to building and evaluating clinician-facing clinical decision support (CDS) systems. Much of the literature treats collaboration quality as a property of the model or a property of trust and transparency mechanisms, while the type of interaction with the model and its outputs receive less attention in examining how collaboration is practically produced during augmented with AI.

\subsection{Human--AI teaming: concepts, shared models, and evaluation}
Human--AI teaming research has developed definitions, conceptual frameworks, and evaluation approaches for human-machine teams. Prior work has emphasized that the definitions of human--AI teaming are often technology-centric, motivating the need for cross-disciplinary human-centered definitions and taxonomies \cite{Berretta2023}. Another body of work reinterprets human--AI collaboration through theories of human teamwork, shifting the focus on constructs, such as \textbf{coordination, interdependence, communication, and team cognition,} as a basis for understanding collaboration \cite{Braun2023,Lou2025}. At a more theoretical level, recent work highlights the importance of \textbf{shared mental models} between AI and humans to facilitate coordination and effective team behavior, extending classic ideas related to human-human teamwork into human--AI teamwork \cite{andrews2023role}.

Another line of work explores how to measure and test human-machine (agent) teams. Testbeds and proposed metrics focus on team effectiveness, team processes, and interaction quality across different task settings \cite{Cavanah2020,Wilkins2024}. Situation awareness has also been adapted into frameworks intended to describe and assess teaming states in human--AI settings \cite{Gao2023}. Additionally, other works outline the requirements for such team-centered AI from a technical perspective, such as cognitive competence, reinforcement learning, and semantic communication \cite{Hagemann2023}. In a multi--agent environment, task allocation strategies for role-based human machine collaboration formalize how responsibilities can be distributed across agents~\cite{Zhang2025}. From a behavioral perspective, researchers emphasize transparency, explainability, and situation awareness as core considerations for supporting human--AI teams \cite{Endsley2023}.

While the literature provides criteria for what meaningful human--AI teaming should look like and how it might be understood and evaluated, much of these involve macro constructs, such as shared mental models, team situation awareness, and team cognition, rather than examining the micro interaction mechanisms through which collaboration unfolds, succeeds, or fails. Our work builds on this gap by characterizing interaction patterns that shape how clinician--AI collaboration emerges under realistic constraints of a high stakes context.

\subsection{Clinical Decision Support Systems: history and evolution}
Clinical Decision Support Systems (CDSS) originated with the administrative and monitoring frameworks of the 1950s and 1960s \cite{gorry1968experience}. This era laid the groundwork for the seminal ``expert systems'' of the 1970s, such as MYCIN and CASNET, which utilized rule-based heuristics to emulate expert diagnostic reasoning \cite{shortliffe1986medical, kong2008clinical}. While these early Computer-Assisted Medical Decision (CMD) systems operated primarily as standalone interactive tools \cite{reggia1981computer}, the field has since evolved from static knowledge representations to sophisticated, data-driven architectures employing Bayesian models and artificial neural networks \cite{musen2021clinical}.
Large language models, owing to their ease of access and interaction, are now a popular choice for seeking clinical decision support both by clinicians and patients. LLMs, especially those adapted for clinical use, such as OpenEvidence~\footnote{\url{https://www.openevidence.com/}}, can perform a range of functions, such as retrieve and summarize medical knowledge, draft differential diagnoses, and suggest diagnostic or treatment options from free text clinical context. Traditional clinical decision support systems are also being combined with large language models to enable natural language interfaces on traditional machine learning models so they can make interaction easier and overall the system potentially more usable at the point of care~\cite{vrdoljak2025review}. However, these systems remain decision aids rather than autonomous decision makers because they can hallucinate, reflect biases,  have inherent privacy and security concerns, and have variable performance by specialty and task~\cite{vrdoljak2025review}. Empirical CDSS work also suggests that adding an LLM layer can improve ease of use and perceived trust, particularly when responses include citations, but clinicians still report practical downsides such as long outputs and limited transparency about training data~\cite{Rajashekar24human}. Finally, recent clinical evaluations show that current LLMs can miss guideline concordance and are sensitive to how information is presented, which creates workflow integration risks and reinforces the need for structured oversight and rigorous validation in CDSS settings\cite{hager2024evaluation}. 

Despite the promise of LLMs for clinical use, recent randomized controlled trials of LLM use by clinicians show mixed evidence on clinical performance. While LLMs did not significantly improve clinical reasoning compared with conventional resources \cite{goh2024large}, it has the potential to improve  management reasoning in complex cases \cite{goh2025gpt}. Such mixed evidence warrants a closer examination of LLM's capabilities and clinician-LLM collaboration. A review of LLM-based applications for patient care shows several design and output based limitations that include limited dialogue capabilities, prompt-dependent output, non-reproducibility of the output, and inability to understand medical information or engage in medical reasoning~\cite{busch2025current}. While there is some understanding of LLM's properties that support or limit use by clinicians,  there is limited understanding of clinicians' interaction with LLMs, how design affects use, and how it compares with other types of interaction, such as traditional user interfaces.

Prior work has also noted limitations in the evaluation of LLM based CDSSs, describing three levels of failures: (1) what gets measured, where studies focus overly on accuracy or exam style QA and under measure deployment relevant outcomes like safety harms, bias and fairness, and real workflow feasibility, leaving results hard to compare and easy to overinterpret; systematic reviews repeatedly note fragmented evaluation methods and limited attention to these dimensions \cite{bedi2025testing}. (2) what gets tested, where models are validated on curated summaries rather than messy real world inputs, so performance drops when confronted with long unstructured notes and incomplete context~\cite{roeschl2025assessing}. (3) how the system behaves in context, where LLM outputs are sensitive to prompt wording and to the quantity and order of information, may hallucinate or produce unreliable recommendations, and can miss guideline concordance, which blocks safe integration into clinical workflows even when performance on benchmarks is strong~\cite{hager2024evaluation}.  Our work builds on this body of work to generate empirical evidence on clinicians' interaction with LLMs and their perception of LLM's behaviors.

\subsection{Human-centered clinical decision support and clinician-facing AI interaction design}
The medical informatics literature explores the role of CDSSs from the perspective of information management, attention management, and patient safety and diagnostic supports \cite{sutton2020overview,musen2021clinical}. Early HCI and clinical work  emphasized interaction design under real constraints, showing how form factors and workflow realities shape CDSS adoption \cite{Anokwa2012}. Early reviews of CDSS evaluation document heterogeneous assessment practices and motivate stronger approaches to evaluating CDSS in context \cite{Bright2012}. Noting clinicians' reluctance to use computational decision-support tools, HCI and design work has also noted the need for making CDSS unremarkable, such that they only draw attention and slow clinicians down when necessary, remaining unobtrusive at other times \cite{yang2019CHICDSS}. 

More recent HCI work on CDSSs focuses explicitly on human--AI collaboration in the context of trust and transparency through explainable AI. Explanations can increase clinicians’ trust in AI recommendations and design can play a key role in avoiding overreliance \cite{Panigutti2023}. Prior research has explored how explanations are related to user trust and reliance, as well as what information users would find helpful to better understand the reliability of a system's decision-making. \cite{Bussone2015}. Prior work argues for transparency as a core requirement for CDSSs and examines how to operationalize transparency and trust in clinical interfaces \cite{Shortliffe2018,Bayor2025}. More work continues this line by examining sociotechnical challenges of medical ML innovation \cite{Hubert2023}, and studying how to make transparency meaningful in specialist practice \cite{Schor2024}, as well as exploring mechanisms to calibrate trust using estimates of human and AI correctness likelihood \cite{Ma2023Trust}. 

Additional work shows that system performance can shape human--AI collaboration in clinical decision making in ways that are not captured by accuracy alone, since clinicians adapt how much they verify, rely on, or discount model output during use \cite{Challen2019}. Other interactive decision support systems close the loop between expert judgment and model behavior by combining learned predictions with domain rules and allowing feature level feedback that aligns the system with expert over time \cite{Lee2021}. Recent work questions the usual approach where the system makes a recommendation and the clinician decides whether to follow it, proposes alternative human--AI interaction models, and demonstrates them through interface prototypes to compare how different configurations influence decision making \cite{Hussain2024}.

These different lines of research establish what human-AI teaming should look like, what LLM-based CDSSs can and cannot do, and how clinicians reason under uncertainty, but they stop short of explaining how clinicians actually use LLMs. What also remains underexplored is how interaction structures shape how clinicians query, interpret, challenge, and act on AI outputs in real clinical workflows. Our work addresses this gap by empirically examining how different interaction modalities and interface designs support or limit collaborative reasoning, particularlywith LLMs.

\section{Method}
\label{sec:method}
We conducted a qualitative study utilizing a think-aloud protocol and semi-structured interview to examine clinical decision-making, understand clinicians’ perceptions on the use of AI and their needs from AI, and obtain feedback on novel designs. The study was conducted in two phases outlined in Table~\ref{tab:study-structure}:

\begin{table*}
\centering
\caption{Study structure across phases and within-session procedure.}
\label{tab:study-structure}
\footnotesize
\setlength{\tabcolsep}{6pt}
\renewcommand{\arraystretch}{1.15}

\begin{tabular}{p{1.5cm} p{5.0cm} p{7.0cm}}
\toprule
\textbf{Phase} & \textbf{Participants} & \textbf{Within-session procedure} \\
\midrule

\textbf{Phase 1} &
\begin{minipage}[t]{\linewidth}\vspace{0pt}
Hospital Medicine (32, M)\\
Academic Hospitalist (30, F)\\
Pediatrics (55, M)\\
Rad-Onc Resident (32, M)\\
Anesthesiology Resident (30, M)\\
Rad-Onc Resident (28, M)\\
CI Fellow / Internist (35, M)
\end{minipage}
&
\begin{minipage}[t]{\linewidth}\vspace{0pt}
Study introduction\\
Initial interview (baseline AI use)\\
Case A task (unguided AI use)\\
Case B task (unguided AI use)\\
Exit interview (extended reflection)
\end{minipage}
\\
\addlinespace[2pt]

\midrule
\textbf{Synthesis} &
\begin{minipage}[t]{\linewidth}\vspace{0pt}

\end{minipage}
&
\begin{minipage}[t]{\linewidth}\vspace{0pt}
Affinity mapping of Phase 1 observations\\
Identification of design dimensions\\
Development of 14 UI variations
\end{minipage}
\\
\addlinespace[2pt]

\midrule

\textbf{Phase 2} &
\begin{minipage}[t]{\linewidth}\vspace{0pt}
Internal medicine hospitalist (31, M)\\
Pediatrician (32, F)\\
General surgeon (32, M)\\
Pediatrician and sports medicine (32, F)\\
Internal medicine (29, M)
\end{minipage}
&
\begin{minipage}[t]{\linewidth}\vspace{0pt}
Study introduction\\
Initial interview\\
Case 1 (guided AI use)\\
UI walkthrough and preference selection\\
Case 2 (unguided AI use)\\
Exit interview (focused)
\end{minipage}
\\
\bottomrule
\end{tabular}
\end{table*}

\textbf{Phase 1: Exploratory Interaction and Requirements Gathering} In Phase 1, we conducted 90-minute sessions with 7 clinicians to map decision behaviors and generate design requirements. Each session began with a 20-minute semi-structured interview covering the participant's specialty, work routines, current AI usage, and broader attitudes toward generative AI in the workplace. Following this, participants reviewed two complex clinical vignettes, each containing multiple tasks developed through a rigorous pipeline involving human review (see Section \ref{sec:case-prep}). We presented these using a progressive disclosure approach: participants first received the clinical scenario, followed by specific tasks sequentially, and finally the decision options (where applicable) as they worked through the problem. This phase concluded with an interview to elicit needs and challenges based on their immediate experience. Additionally, in Phase 1, we used the tasks and the final interview to identify breakdowns in chat based decision support and translate them into interface requirements through affinity mapping. After participants completed the tasks, we asked focused follow up questions to clarify two themes that repeatedly surfaced in their reasoning and feedback: how they wanted uncertainty and alternatives to be expressed, and how much interaction a representation should require during time constrained clinical work. We then synthesized their responses into 14 candidate UI variations that varied these two themes, numerical and non numerical (\textit{numericality}) representations as well as static and interactive formats (\textit{interactivity}).

\textbf{Phase 2: Structured Collaboration and Artifact Review} Phase 2 involved sessions with 5 clinicians and focused on exploring the designs variations based on insights from Phase 1. In this phase, while the tasks were the same as Phase 1, the execution was closely guided. Instead of using AI as and when they would like to, participants were given step-wise instructions in Case A Task 2 and 3 that first involved writing a note to the specialist, and then use that note as a prompt to AI. In task 2, participant were encouraged (but not required) to employ the same procedure for creating their future prompts in that session. This shift aimed to change the role of AI from an open ended source of information to a more specific role and to anchor prompt creation in the act of writing a note for the specialist, a more familiar tasks for clinicians. Furthermore, the design probe session was formalized into a systematic walkthrough of 14 UI variations to explore visual hierarchy, interactivity, and the representation of clinical logic.

All sessions took place online via Zoom and lasted approximately 90 minutes. Prior to the session, participants completed a consent form and three brief questionnaires: the Cognitive Style Indicator (CoSI) \cite{Cools2007}, the Rational and Intuitive Decision Styles Scale \cite{Hamilton2016}, and the Maximization Scale \cite{schwartz2002maximizing}. We collected CoSI to capture differences in how clinicians prefer to structure information, and Rational and Intuitive scales to capture decision preferences that could affect how participants interpret and use model output. Maximization Scale was collected to capture differences in alternative search and decision difficulty that could shape whether clinicians used the LLM for confirmation or exploration. In our sample, participants were almost entirely classified as Rational, and CoSI varied across Planning, Creating, and Knowing, as shown in Table \ref{tab:participant_styles}. Aside from maximization, these measures did not show a clear association with interaction patterns or satisfaction in this study. 

\subsection{Participants}
We recruited US clinicians with a minimum of one year of clinical experience. Recruitment proceeded on a rolling basis, with data collection continuing in each phase until data saturation was reached. This resulted in a total of 12 participants, with 7 included in Phase 1 and 5 in Phase 2. The cohort was predominantly male (n=9), with a mean age of 33 years (range: 28–55). Participants represented a diverse range of clinical specialties, with the largest group specializing in Internal Medicine (n=5). Other represented specialties included Pediatrics (n=3), Radiation Oncology (n=2), Anesthesiology (n=1), and Minimally Invasive Surgery (n=1). The sample equally represented attending physicians (n=6) and trainees (n=6), the latter comprising residents and clinical fellows. Participants were compensated with a \$100 honorarium. The study was approved by the university's Institutional Review Board (IRB).

\subsection{Clinical Vignette and Task Preparation}
\label{sec:case-prep}

We designed the clinical vignettes and associated tasks to reflect real decision-making conditions rather than benchmark-style evaluation settings. This choice was motivated by growing evidence that most evaluations of large language models in health care rely on synthetic tasks, curated prompts, or exam-style questions, and rarely involve real patient care data or authentic clinical workflows. A recent systematic review of 519 studies evaluating health care applications of large language models found that only 5\% used real patient data, with most evaluations focusing narrowly on accuracy while underrepresenting complex decision tasks, uncertainty management, and clinical context~\cite{bedi2025testing}.

\begin{figure*}[!h]
    \centering
    \includegraphics[width=1.0\linewidth, alt="A horizontal four-stage process diagram illustrating the construction of clinical vignettes. It moves from human-led de-identification and filtering of real patient data to human-AI collaborative stages for case selection and abstraction. The process finishes with a human-led validation of final decision prompts and constraints."]{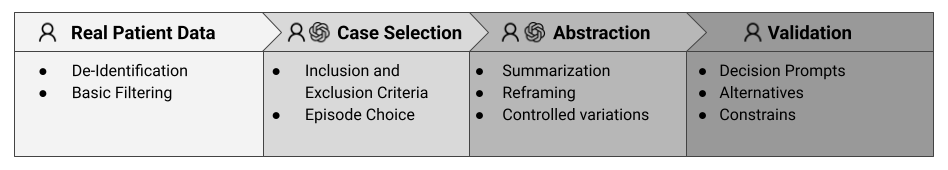}
    \caption{Overview of the patient case, clinical vignette and clinical task construction process. Real patient data serve as the starting point, from which experts select cases followed by specific clinical encounters for those vignettes based on pre-defined inclusion criteria. AI is then used to abstract and reframe selected patient vignette and corresponding encounters under controlled constraints, supporting summarization and limited variation without introducing new clinical content. Final decision-making tasks are constructed and validated by experts to preserve clinical realism.}
    \label{fig:caseprep}
\end{figure*}

In response to these limitations, we adopted a human-in-the-loop pipeline (Figure \ref{fig:caseprep}) grounded in de-identified electronic health record data from real patients. Candidate vignettes were first identified using expert defined criteria through ICD-10 codes, followed by manual review to select patient vignette. Each vignette comprised of multiple clinical \textit{encounter}s. Once patient vignette were selected, specific encounters that exhibited clinical ambiguity, competing decision options, and non-trivial trade-offs were selected for clinical case creation. Identifiable information was removed in compliance with institutional IRB requirements prior to any processing.

For each selected patient case and the corresponding clinical encounter from that case, we retrieved longitudinal context, including clinical notes, laboratory trends, vitals, and medication histories spanning multiple days. These materials were used to generate comprehensive clinical cases via a human-in-the-loop pipeline, after which two domain experts manually reviewed each case for factual correctness, clinical plausibility, and representativeness of real-world decision-making conditions. Tasks were then constructed using a clinical task taxonomy informed by prior work and refined through expert review to align the task with the underlying case structure~\cite{ofstad2016medical,yazdani2017models}.

The cases were designed to surface the kinds of uncertainty, partial information, and competing priorities that shape clinical reasoning in practice. We believe that this grounding was essential for examining how interaction structures and interface representations influence clinician–AI collaboration under realistic constraints.

\subsubsection{Clinical Cases} 
We utilized two clinically grounded vignettes to anchor the study, each representing a distinct decision context while sharing comparable levels of complexity, uncertainty, and time pressure. In this paper, a \emph{Case} refers to a curated snapshot of a real inpatient vignette and a series of curated task that preserves the clinical decision structure, relevant constraints, and competing considerations faced by clinicians in practice, rather than a fully exhaustive patient record. Both cases were derived from real patient data and selected to require tradeoffs across multiple clinical dimensions, making them suitable for examining how interaction designs shape collaborative reasoning with AI.

The first case (Figure \ref{fig:case_a}) involves a 44-year-old woman with cystic fibrosis and a prior bilateral lung transplant who was admitted with sepsis and acute-on-chronic respiratory failure due to pneumonia. The core decision centers on antimicrobial management in the context of chronic immunosuppression, renal dysfunction, and a severe penicillin allergy. Multiple viable treatment options exist, each differing in efficacy, renal safety, guideline alignment, and interaction with immunosuppressive therapy. This case foregrounds a relatively focused decision with well-defined alternatives, allowing us study how clinician approach a well defined but complex decision.

The second case (Figure \ref{fig:case_b}) concerns a 67-year-old kidney transplant recipient presenting with septic shock, acute pulmonary edema, and acute kidney injury superimposed on complex chronic comorbidities, including heart failure and chronic infected pressure ulcers. Unlike the first case, this scenario involves urgent, multi-criteria decision processes and heightened physiological instability, with antimicrobial selection intertwined with fluid management, renal preservation, and immunosuppression. As a result, the decision space is broader and less easily reducible to a single optimization problem. 

\subsubsection{Task Design}

For each clinical vignette, we defined a set of tasks intended to elicit clinician reasoning under realistic decision-making conditions. In this paper, a \emph{task} refers to a structured prompt that asks participants to articulate a non--trivial clinical decision, justify their reasoning, or revise their plan in response to new information. Tasks were designed to surface how clinicians interpret evidence, prioritize constraints, and interact with AI to provide input and assess generated recommendations, rather than to test factual recall or guideline knowledge.

Tasks differed across the two vignettes to reflect differences in clinical structure and reasoning demands (see table \ref{tab:appendix_tasks}). Case~A employed a \emph{guided task} structure, in which participants were prompted to consider a specific decision point, evaluate a constrained set of alternatives, and reflect on how additional perspectives—including AI-generated specialist input—might influence their judgment. The guided format was used to support focused comparison among options in a clinically well-bounded decision space. In Case~B, tasks were open-ended and unguided, requiring participants to independently synthesize patient information, interpret evolving data, and determine next steps without explicit scaffolding. These tasks emphasized exploratory reasoning and allowed participants to define their own decision frames. We utilized this distinction to examine interaction design across different types of tasks, which may result in different reasoning modes. Unguided tasks foreground how clinicians initiate and structure decisions, while guided tasks foreground how they compare alternatives, engage with counterarguments, and revise plans.

{Data Analysis and AI-Assisted Coding}
Our analysis combined structured coding with qualitative thematic analysis. The structured coding used an explicit analytic framework to characterize how clinicians and the AI coordinated work during task execution, including how information was requested, interpreted, contested, and incorporated, while remaining grounded in observable transcript evidence.

\textbf{Units of analysis.}
All session recordings were transcribed and de-identified prior to analysis. Each of the 12 sessions was segmented into four units corresponding to the initial interview, clinical Case A (Vancomycin replacement and consultation workflow), clinical Case B (emergency department intervention and antibiotic update), and the concluding interview or UI design walkthrough. Structured coding assisted by AI was only applied to the Case A and Case B segments, where each case was examined strictly using a pre--defined code book.   

\textbf{Coding schema development.}
To analyze clinical case A and B, the coding schema was derived iteratively from our primary question described in \ref{tab:research-questions}. The final schema consisted of 15 discrete variables grouped across three analytic dimensions focusing on subquestions for analysis (1) role positioning and role shifts, (2) construction and breakdown of shared understanding over multiple turns, and (3) evaluation and use of model outputs. Variables captured observable interaction properties such as when model use occurred during the task, the role assigned to the model (e.g., information lookup versus option generation), query intent, amount of case context provided, reading strategy (full reading versus selective scanning), visible comprehension issues, follow-up behavior, and whether model output influenced task decisions. Each variable was defined with mutually exclusive categories and explicit decision rules to minimize interpretive ambiguity.

\begin{figure}
    \centering
    \includegraphics[width=0.9\linewidth, alt="Flowchart illustrates a multi-agent analysis pipeline where an AI assistant processes session transcripts through five parallel AI judges. The judges select questions from a bank to generate classifications and evidence\, which are then reconciled through a manual check by a researcher to produce a final coding decision."]{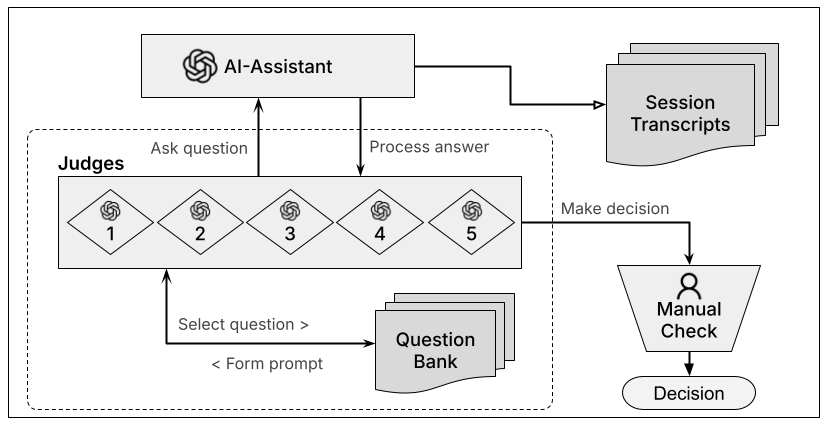}
    \caption{Session transcripts were segmented by task encounter and analyzed using a structured coding schema. For each coding question, five independent AI agents were prompted in parallel using the same transcript segment and schema definitions. Each agent produced a classification along with verbatim evidence quotes. These independent outputs were then reviewed through a manual check, where the researcher reconciled disagreements and determined the final coding decision used in analysis.}
    \label{fig:coding}
\end{figure}

\textbf{AI-assisted structured coding.}
To apply the coding schema across all transcripts while staying grounded in transcript evidence, we implemented an automated pipeline using the OpenAI Assistants API (see Table \ref{tab:model_param}) and a custom script (see Figure \ref{fig:coding}). For each (session, case) pair, we ran five independent coding passes in separate assistant threads. Each thread was initialized with the full coding guide and schema, then received the same fixed sequence of 15 coding questions. For every question, the assistant was instructed to select exactly one category and provide verbatim evidence quotes from the transcript to justify the classification, or return ``Not found in transcript'' when the relevant information was absent. This produced five parallel classifications per question for each encounter. The pipeline logged all assistant outputs, evidence quotes, and run metadata to a structured JSONL file for downstream aggregation and auditing. We built a simple transcript coding viewer (Figure \ref{fig:coding_ui}) to inspect and adjudicate the five parallel agent outputs. The interface lists the 15 coding questions as rows and the five agent ``judges'' as columns, with per cell links to view the extracted evidence quotes, and navigation controls to move across sessions and vignettes.

\begin{table*}[t]
\centering
\caption{Research questions and corresponding coded variables.}
\label{tab:research-questions}
\footnotesize
\setlength{\tabcolsep}{6pt}
\renewcommand{\arraystretch}{1.15}

\begin{tabular}{p{2.2cm} p{10.8cm}}
\toprule
 & \textbf{Question / Coded variables} \\
\midrule

\textbf{Main Question} &
How do clinicians coordinate work with a large language model during structured clinical tasks, and what interaction patterns underlie use of the LLM? \\
\midrule

\textbf{Sub Question 1} &
How do clinicians position the model’s role during tasks, and when does that role shift between instruction following and contributing to task reasoning? \\ \\
& Q1: SQ1-1: Did the clinician use the model in this task encounter? (no, yes) \\
& Q2: SQ1-2: When did model use occur relative to the task? (start, middle, end, multiple points) \\
& Q3: SQ1-3: What role was the model used for? (information lookup, option generation, critique, drafting, other) \\
& Q4: SQ1-4: What was the clinician’s query intent? (diagnosis or differential, guideline or evidence, dosing or safety, test interpretation, documentation or phrasing, patient communication, workflow, other) \\
& Q5: SQ1-5: What was the clinician’s query? (verbatim) \\
\midrule

\textbf{Sub Question 2} &
How do clinicians and the model build a shared understanding of the case and the task, and manage missing context, uncertainty, and disagreement? \\ \\
& Q6: SQ2-1: How much case context was included in the initial query? (none, some, detailed) \\
& Q7: SQ2-2: Did the clinician add or correct case details after the response? (no, yes) \\
& Q8: SQ2-3: How did the clinician read the response? (full read, selective scan) \\
& Q9: SQ2-4: Were there visible comprehension issues while using the response? (no, yes) \\
& Q10: SQ2-5: Did the clinician reframe the query after the response? (no, yes) \\
\midrule

\textbf{Sub Question 3} &
How do clinicians evaluate and leverage model outputs, including verification, follow-up questioning, and decisions to adopt, modify, or disregard the model’s input? \\ \\
& Q11: SQ3-1: If the query was reframed, what changed in the query? (added context, narrowed scope, requested structure, requested evidence, changed goal, other) \\
& Q12: SQ3-2: Did the clinician judge the response as correct? (yes, no, unclear) \\
& Q13: SQ3-3: Was the clinician satisfied with the response? (satisfied, partly satisfied, not satisfied) \\
& Q14: SQ3-4: If not satisfied, what did the clinician do next? (re-prompt, ask for evidence, ignore and proceed, switch resource, stop using model, other) \\
& Q15: SQ3-5: Did the clinician ask a follow-up question, or change their task answer or plan after reading the response? (no, follow-up only, change only, both) \\
\bottomrule
\end{tabular}
\end{table*}

\textbf{Aggregation and validation.}
For the primary quantitative results, we aggregated the five independent agent outputs per question to identify dominant interaction patterns across both phases and clinical vignettes. Disagreements were retained rather than collapsed prematurely, allowing us to characterize uncertainty and ambiguity in observed behaviors.

To validate and complement the automated coding, the lead researcher manually coded the full dataset using the same schema and decision rules before coding with AI. These manual codes were reviewed alongside the five agent outputs and their associated evidence quotes for each data point using our coding user interface displayed in Figure~\ref{fig:coding_ui}. This comparison step was used to identify inconsistencies, surface ambiguous transcript segments, and verify that final classifications were defensible based on observable evidence and aligned with research questions. The researchers determined the final label for each variable after reviewing both manual and AI-generated labels. Table \ref{tab:agreement_summary} displays the inter--agreement between the agents and manual coding.

\begin{table*}[!ht]
\centering
\caption{The table reports two reliability metrics across the coding schema: (1) \textit{Agent Agreement}, representing the proportion of encounters where all five independent agents reached unanimous consensus on the label; and (2) \textit{Human-AI Agreement}, representing the concordance between the aggregated AI consensus (majority vote) and the expert human coder's ground truth. Dashes (--) indicate fields containing verbatim text or open-ended descriptions where categorical agreement was not applicable.}
\label{tab:agreement_summary}
\begin{tabular}{lllcc}

\hline

\textbf{ID} & \textbf{Question Topic} & \textbf{Agent Agreement (\%)} & \textbf{Fleiss' Kappa (agents)} & \textbf{Human-AI Agreement (\%)} \\
\hline
Q1 & Did clinician use model? & 94.1 & 0.864 & 100.0 \\
Q2 & Timing of use? & 58.8 & 0.658 & 79.2 \\
Q3 & Role of model? & 64.7 & 0.582 & 91.7 \\
Q4 & Query Intent? & 52.9 & 0.615 &87.5 \\
Q5 & Verbatim Query? & 88.2 & --- & -- \\
Q6 & Context included? & 64.7 & 0.493 & 91.7 \\
Q7 & Added details after? & 94.1 & 0.781 & 100.0 \\
Q8 & How read response? & 70.6 & 0.327 & 95.8 \\
Q9 & Comprehension issues? & 94.1 & 0.556 & 95.8 \\
Q10 & Reframed query? & 100.0 & 0.724 & 100.0 \\
Q11 & What changed? & 100.0 & 0.689 & -- \\
Q12 & Judged correct? & 73.3 & 0.312 &95.8 \\
Q13 & Satisfied? & 64.3 & 0.447 &95.8 \\
Q14 & Next action? & 78.6 & 0.518 & 75.0 \\
Q15 & Follow-up/Change? &  85.7 & 0.763 & -- \\
\hline
\end{tabular}
\end{table*}

\textbf{Thematic analysis.}
In parallel, we conducted an inductive thematic analysis of the semi-structured interview segments and the Phase 2 UI design walkthroughs. These portions were not subjected to structured coding using the coding schema but were analyzed to capture participant and AI reasoning, preferences, and critiques related to interaction structure, information hierarchy, and cognitive effort. Feedback on UI variants and speculative design prompts were aggregated to derive qualitative insights that contextualize the interaction patterns.

\begin{figure*}
    \centering
    \includegraphics[width=0.60\linewidth, alt="A medical dashboard displays four antibiotic options as vertical cards. Each card features a prominent numerical score\, such as 99 for Linezolid\, alongside color-coded indicators for factors like kidney impact and lung penetration. A strategy panel on the left allows users to toggle between safety-focused or coverage-focused recommendations\, while a bottom panel provides a text-based rationale for the selected drug."]{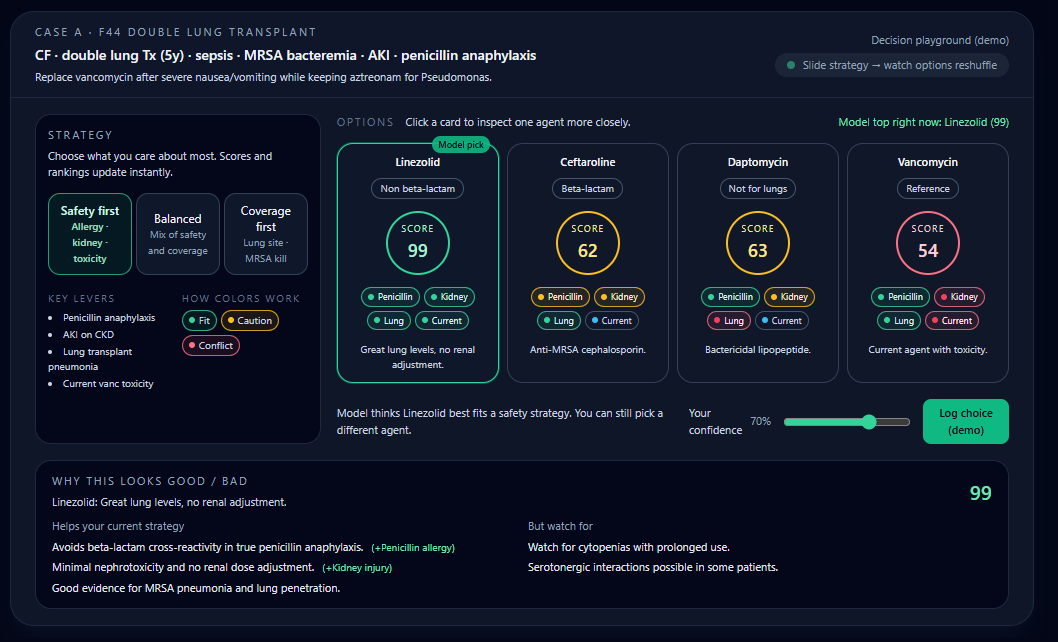}
    \caption{UI Variation: \textbf{Scores}. Interface highlighting numerical scores and discrete decision factors. The dashboard allows clinicians to toggle between decision strategies, such as ``Safety first'' or ``Coverage first,'' while presenting AI-generated match scores and color-coded rationale for different antibiotic options.}
    \label{fig:scores}
\end{figure*}

\begin{figure*}
    \centering
    \includegraphics[width=0.60\linewidth, alt="A decision tree  illustrates the logic for choosing a new antibiotic by filtering three candidates through clinical checkpoints. The visualization shows one option eliminated for ineffective lung penetration and another labeled conditional due to renal safety requirements. The optimal recommendation is highlighted at the end of a successful logical path."]{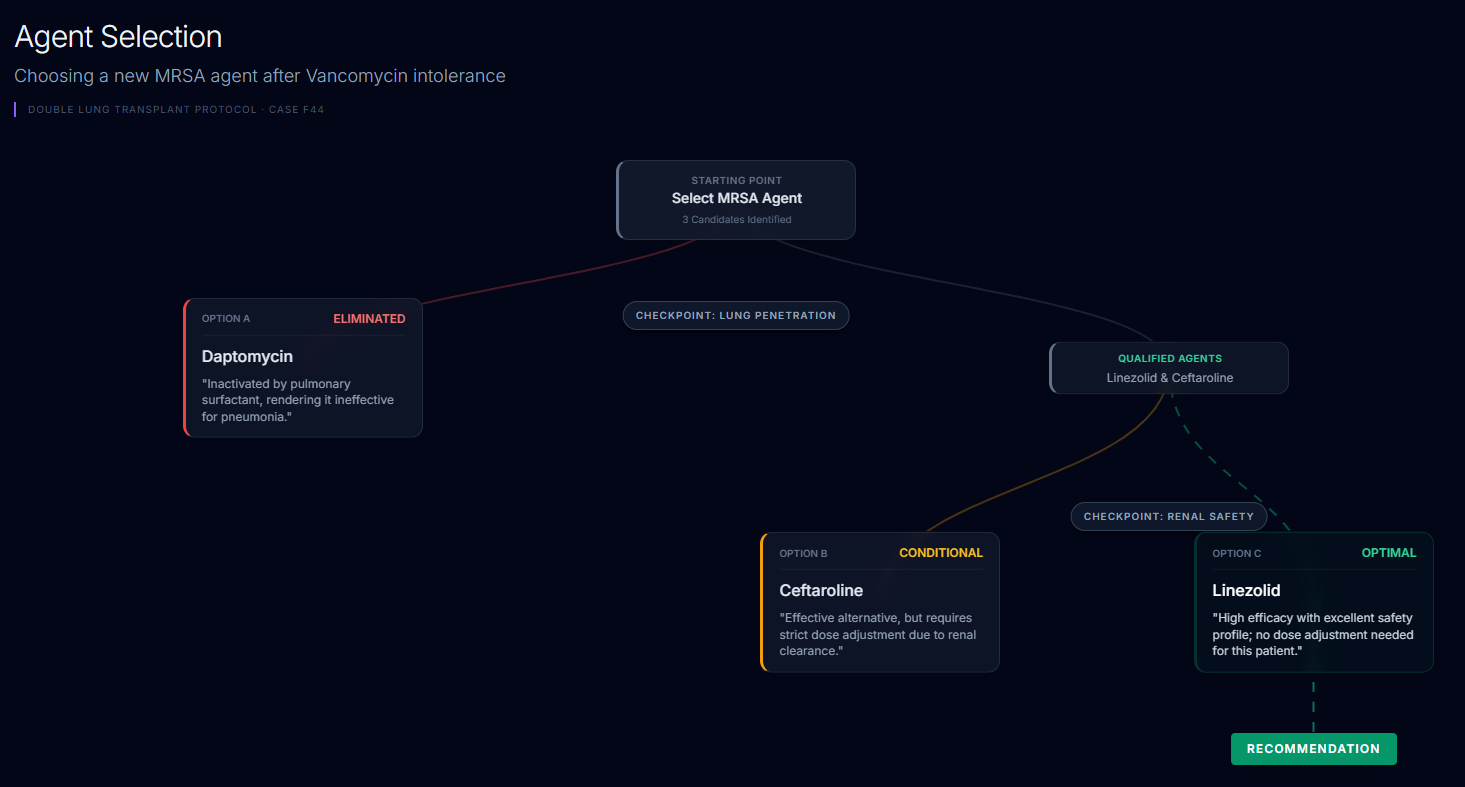}
    \caption{UI variation: \textbf{Graph}. This format externalizes clinical logic by visualizing the ``pathway'' of a decision, showing why certain options were ruled out or remained as viable candidates based on specific clinical checkpoints like lung penetration and renal safety.}
    \label{fig:graph}
\end{figure*}

\begin{figure*}
    \centering
    \includegraphics[width=0.60\linewidth, alt="Three vertical cards titled Protocol Selection compare clinical data for the antibiotics Ceftaroline\, Linezolid\, and Daptomycin. Each card displays structured information on safety scores\, dosage adjustments\, lung penetration\, and clinical guidelines. Linezolid is highlighted as the recommended choice with high renal safety and optimal lung coverage\, whereas Daptomycin displays a red warning for lung inactivation."]{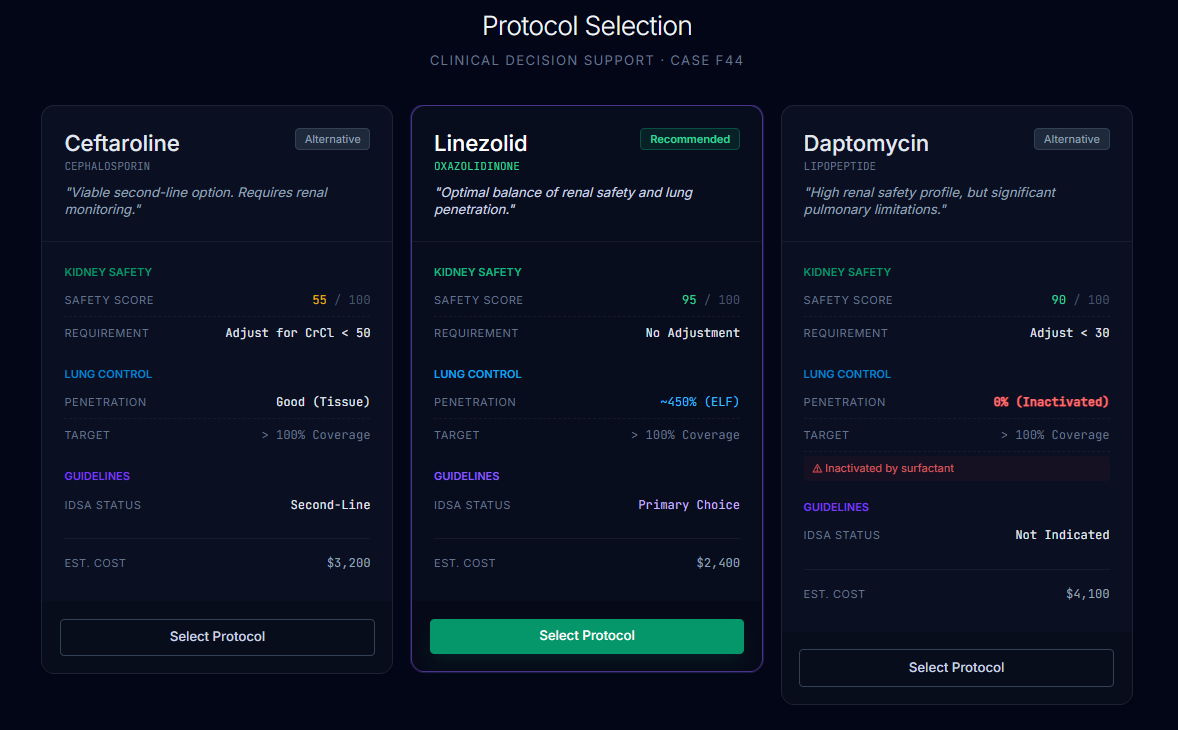}
    \caption{UI variation: \textbf{Cards}. This design organizes clinical data points—such as renal safety, lung penetration, and IDSA status into side-by-side cards to support rapid comparative reasoning and identifying a leading clinical option.}
    \label{fig:cards}
\end{figure*}

\section{Results}
This section reports the interaction patterns observed across the design walkthrough sessions in our study. We examine how clinicians positioned the model’s role, how they attended to and evaluated its outputs, and how these patterns changed when the interaction setup and process were controlled within the sessions. The results also describe differences associated with individual decision and cognitive styles, as well as clinicians’ responses to alternative visual representations of the same underlying recommendations.

\subsection{LLMs were primarily used as information lookup tools rather than as partners in clinical reasoning}
Clinicians consistently positioned the model as a tool for retrieving factual clinical information rather than as a partner in diagnostic or therapeutic reasoning. Across the sessions, this positioning of the model's role was evident in both how participants described the system and how they interacted with it during task execution. In $86$ percent of observed encounters, clinicians used the model to fetch guidelines, dosing protocols, or contraindications instead of engaging it in deliberation about diagnoses or treatment plans. Participants explicitly articulated this boundary. One hospitalist described the system as \textit{``basically like UpToDate\footnote{UpToDate (uptodate.com) is a searchable medical database that healthcare professionals use to find evidence-based information for diagnosing and treating patients.} but faster,''} explaining that they were \textit{``just checking dosing and contraindications, not asking it to make the decision''} (P2). A radiation oncology resident similarly framed the model as \textit{``a smart medical student''} that could \textit{``pull facts quickly''} but should not be trusted with clinical judgment (P4).  

Once the desired information was obtained, clinicians routinely stopped engaging with the model output and discarded the rest of the response, even when it contained clinical reasoning or alternative options. This \toollike use of the model was visible in the interaction sequences. For example, P1  first verbalized a complete antibiotic plan aloud and then queried the model specifically for vancomycin renal dosing. When the model returned a multi-paragraph response that included rationale and alternative treatments, the participant visually scanned the text, stopped reading once the dosing information appeared, and dismissed the remaining content as unnecessary stating \textit{``“There’s maybe just a couple words that are kind of helpful.''}. Comparable patterns were seen across the sessions. Participants typically terminated engagement as soon as the targeted information was located, and the additional explanations or alternatives generated by the model rarely prompted deeper thinking, reconsideration, or modification of their original plans.

The way clinicians formulated their queries further reinforced the model’s role as an information lookup tool rather than a reasoning partner. Participant inputs were narrow and factual, such as \textit{``recommended vancomycin dosing adjustments in CKD''}(P3), ``\textit{penicillin-allergy alternatives for MRSA pneumonia''}(P6),  or `\textit{`contraindications and side effects of linezolid versus daptomycin''} (P5). These queries did not invite the model to weigh trade-offs, challenge assumptions, or propose a treatment strategy. Even when the model provided extended reasoning, participants seldom engaged with it or followed up on it, treating the output as a source of isolated facts rather than a basis for collaborative clinical reasoning. Participants were aware that the model was capable of deeper analysis, even when they chose not to use it that way. This was expressed through how clinicians talked about constraining the model’s role. P3 noted, \textit{``It clearly knows more than it’s showing me here… it’s capable of doing deeper analysis.''} P7 similarly said, \textit{``If I wanted it to really reason this out, I think it could.''} This restraint was framed as a way to stay in control and limit output. P10 explained, \textit{``I’m deliberately keeping it focused, because otherwise it gives me too much,''} and P4 added, \textit{``I don’t want it driving the reasoning unless I ask for it.''}.

Clinicians’ reading behavior also reflected the model’s role as an information lookup tool. Despite providing rich case context in $82$ percent of encounters, participants rarely read the model’s reasoning in full. In $70$ percent of observed interaction encounters, they instead used selective scanning to locate specific keywords, numbers, or validation signals. One hospitalist stated, \textit{``I’m not actually reading all of this, I’m just looking for the part that tells me if what I’m doing is okay''} (P1). Another participant said, \textit{``'I’m skimming until I see the number or the drug name I care about,''} adding that the surrounding explanation was \textit{``nice but not necessary''} (P2). In multiple sessions, clinicians verbally disengaged while the model output was still visible, using phrases such as \textit{``that’s enough,''} \textit{``I got what I need,''} or \textit{``the rest is just explanation''} (P1, P4, P7).

These strategies were also visible in how participants interacted with the model on screen. Observed interaction behavior showed the same selective scanning and early disengagement that participants described. In P1’s session, after submitting a detailed prompt describing patient history and constraints, the participant scrolled rapidly through the model’s response, paused briefly when encountering a dosing table, and then immediately shifted visual attention back to the case description without reading the remainder of the text. In P3’s session, the participant skimmed the first few lines of the output, searched within the response for a specific medication name, and then proceeded to answer the task question without returning to the model output again. 

\subsection{Positioning the LLM in a familiar role and scaffolding prompt creation triggered engagement beyond information lookup}

Clinicians engaged more critically with the model in Phase 2 when the task required them to first write down a case summary and a decision question that would be used as the model prompt (as compared to free form prompting). When asked to include the patient case information as a part of initial prompt alongside the note to specialist, P7 mentioned: \textit{``Yeah... That would be much better, ... so you can get a better response''} and confirming that this is a more fitting comparison to what an actual consultant would have access to in practice.  Participants checked whether the model had picked up the main problem and the relevant constraints from their summary. P8, for example, called out whether the response reflected the main concern and marked which parts of the output matched the constraints of the case. P12 described starting by extracting the \textit{``top three recommendations''} and then going to a specific section of the response to look more closely at details. This setup also led clinicians to comment on how the model organized its advice. P11 evaluated whether the suggested steps matched their clinical sequencing and pointed out when the model’s ordering did not fit their framing, saying that a \textit{``hemodynamic assessment… isn’t really, like, a step''} and that the model \textit{``should have said, like, put oxygen on his face''}. P12 described a deliberate reading strategy, explaining that they would \textit{``look at the top three recommendations first''} and then \textit{``go to hemodynamic management''} to inspect the details. Differences between Phase 1 and Phase 2 are described in \ref{tab:phase_setup}. Positioning the model as a specialist, creating a message for a specialty consult (an artifact they are used to creating in practice) and using it as a prompt for the model resulted in engagement beyond information lookup. This time, clinicians checked if the model understood what they wanted to convey and paid close attention to its recommendations.

\begin{table*}[t]
\centering
\caption{Differences in task setup between Phase 1 and Phase 2}
\label{tab:phase_setup}
\begin{tabular}{p{6cm} p{6cm}}
\toprule
\textbf{Phase 1} & \textbf{Phase 2 (Specialist Note)} \\
\midrule
Participants could ask the model anything at any time while working on the case. & Participants were required to first write a Specialist Note summarizing the case and their question before using the model. \\
\addlinespace
Prompts were typed directly into the chat and were typically short and task focused. & The Specialist Note was used as the model prompt, producing a longer and more structured input. \\
\addlinespace
The model was used as a general information source. & The model was positioned as a specialist responding to a clinician’s formal consult note. \\
\bottomrule
\end{tabular}
\end{table*}

\subsection{Different decision styles led to different uses of the LLM}
\begin{figure*}
    \centering
    \includegraphics[width=0.90\linewidth, alt="Three side-by-side boxplots compare maximizers and satisficers on High Standards\, Alternative Search\, and Decision Difficulty subscales. While High Standards scores are similar for both groups\, maximizers show higher levels of Alternative Search and significantly greater Decision Difficulty compared to satisficers."]{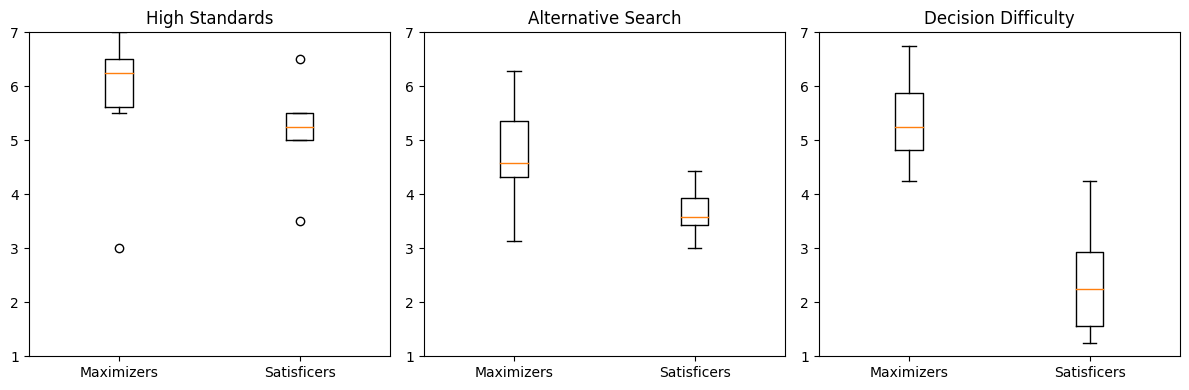}
    \caption{Boxplots showing Maximization Scale subscale scores for maximizers and satisficers in the study sample. The three panels report High Standards, Alternative Search, and Decision Difficulty. High Standards are similar across the two groups, while Alternative Search is higher among maximizers than satisficers, and Decision Difficulty shows the largest separation, with maximizers reporting greater difficulty committing to a decision.}
    \label{fig:sat_max}
\end{figure*}

Participants were grouped as maximizers or satisficers based on their scores on the Maximization Scale (MS)~\cite{schwartz2002maximizing}. The scale captures three components of decision style: High Standards, Alternative Search, and Decision Difficulty. In our sample (Figure~\ref{fig:sat_max}), High Standards did not meaningfully separate the two groups, while Alternative Search showed a moderate separation and Decision Difficulty showed a strong separation.

We observed differences in how clinicians in each group used the model. Maximizers tended to use the LLM to validate decisions they had already thought through rather than developing or revising those decisions. Their prompts were narrow and confirmatory, and their engagement with the output typically stopped once a validating cue was found. P1, for example, stated,\textit{ ``I’m not actually reading all of this, I’m just looking for the part that tells me if what I’m doing is okay,''} before taking attention away from the remainder of the response. Other maximizers similarly treated the model as a source of isolated confirmation signals rather than a material to be worked through. P10 described the useful part of the output as ``an actual justification for the selection,''`` indicating that the rest of the response was not central to their decision process. Two maximizers also resisted adopting the model’s framing or structure of the output when it did not match their own. P11 rejected the model’s stepwise organization by saying, \textit{``I already did this kind of assessment… this isn’t really, like, a step,''} and P4 questioned the authority of model-generated guidance altogether, asking, \textit{``What are these sources of guidance that make random connections? Sure, because I’m gonna believe that!''}

\begin{table*}
\centering
\caption{Participant Classification by Cognitive, Decision, and Maximization Styles}
\begin{tabular}{l|ccc|cc|cc}
\toprule
 & \multicolumn{3}{c|}{\textbf{Cognitive Style (CoSI) } \cite{Cools2007} } & \multicolumn{2}{c|}{\textbf{Decision Style}\cite{Hamilton2016} }  & \multicolumn{2}{c}{\textbf{Maximization} \cite{schwartz2002maximizing}} \\
\textbf{Session} & \textbf{Plan} & \textbf{Create} & \textbf{Know} & \textbf{Rational} & \textbf{Intuitive} & \textbf{Maximizer} & \textbf{Satisficer} \\
\midrule
P1 & $\checkmark$ &  &  & $\checkmark$ &  & $\checkmark$ &  \\
P2 &  &  & $\checkmark$ & $\checkmark$ &  &  & $\checkmark$ \\
P3 &  & $\checkmark$ &  &  & $\checkmark$ &  & $\checkmark$ \\
P4 & $\checkmark$ &  &  & $\checkmark$ &  & $\checkmark$ &  \\
P5 &  & $\checkmark$ &  & $\checkmark$ &  &  & $\checkmark$ \\
P6 & $\checkmark$ &  &  & $\checkmark$ &  & $\checkmark$ &  \\
P7 & $\checkmark$ &  &  & $\checkmark$ &  &  & $\checkmark$ \\
P8 &  & $\checkmark$ &  & $\checkmark$ &  & $\checkmark$ &  \\
p9 &  & $\checkmark$ &  & $\checkmark$ &  &  & $\checkmark$ \\
p10 & $\checkmark$ &  &  & $\checkmark$ &  & $\checkmark$ &  \\
p11 & $\checkmark$ &  &  & $\checkmark$ &  & $\checkmark$ &  \\
p12 &  &  & $\checkmark$ & $\checkmark$ &  &  & $\checkmark$ \\
\bottomrule
\end{tabular}

\label{tab:participant_styles}
\end{table*}

Satisficers, in contrast, were more likely to use the model output as part of their ongoing reasoning rather than as a final check. Their prompts were more open-ended and often aimed at comparing options or resolving uncertainty. P5, for instance, said, \textit{``I’m not sure which one has less GI side effects, and so I think that would guide me''}, using the model to weigh alternatives rather than to confirm a fixed choice. Satisficers also engaged more actively with the content of the responses, including critiquing and reframing them when needed. P2 challenged the model’s characterization of a case by stating, \textit{``I would call it something different… the emphasis should be on his respiratory status,''} and P7 used the model to structure information for another agent to reason about, explaining that the goal was \textit{``so that they could… make their own plan, and then call me back''}. P12 described the model output as something they would \textit{``probably read, if I was unsure about this case,''} treating it as a reference for thinking rather than as a quick lookup.

\begin{table*}[t]
\centering
\caption{Maximizers vs. Satisficers: Evidence from in-task LLM use}
\label{tab:satmax_quotes}
\begin{tabular}{p{2.2cm} p{1.2cm} p{4.2cm} p{6.8cm}}
\toprule
\textbf{Group} & \textbf{P\#} & \textbf{Observed behavior} & \textbf{Verbatim evidence} \\
\midrule

\multirow{5}{*}{Maximizer}
& P1  & Looked for validation and stopped reading  
& \textit{``I’m not actually reading all of this, I’m just looking for the part that tells me if what I’m doing is okay.''} \\

& P4  & Rejected model authority  
& \textit{``What are these sources of guidance that make random connections? Sure, because I’m gonna believe that.''} \\

& P8  & Treated output as unstructured  
& \textit{``It just spits out, like, whatever it wants.''} \\

& P10 & Extracted one justification cue  
& \textit{``What’s nice in the answer here is that it kind of gives me an actual justification for the selection...''} \\

& P11 & Rejected model’s step structure  
& \textit{``I already did this kind of assessment... this isn’t really, like, a step.''} \\

\midrule

\multirow{5}{*}{Satisficer}
& P2  & Critiqued and reframed output  
& \textit{``I would call it something different... the emphasis should be on his respiratory status.''} \\

& P5  & Used model to resolve uncertainty  
& \textit{``I’m not sure which one has less GI side effects, and so I think that would guide me.''} \\

& P7  & Structured output for another agent  
& \textit{``So that they could have enough pertinent information to look this patient up, make their own plan, and then call me back.''} \\

& P9  & Asked model to compare options  
& \textit{``...the one with less side effects...''} \\

& P12 & Read output as a reference  
& \textit{``This is something that I would probably read, if I was unsure about this case.''} \\

\bottomrule
\end{tabular}
\end{table*}

Within the same \toollike interaction setting, maximizers and satisficers engaged with the model differently. Maximizers tended to keep the framing and synthesis in their own hands and used the LLM narrowly for confirmation, while satisficers were more willing to integrate the model output into their reasoning, using it to explore alternatives, critique assumptions, and adjust their understanding of the case.

\subsubsection{Cognitive and Decision Style Profiles}
On the Rational-Intuitive Decision Styles Scale, the cohort was almost entirely classified as \textit{Rational} ($n=11$), with only one participant identified as \textit{Intuitive}. The Cognitive Style Indicator showed a broader distribution, with half the cohort exhibiting a \textit{Planning} style ($n=6$), followed by the \textit{Creating} ($n=4$) and \textit{Knowing} ($n=2$) styles. Individual classifications for all participants are detailed in Table~\ref{tab:participant_styles}. Unlike the maximization trait, however, these measures did not strongly differentiate participants' interaction patterns or satisfaction levels in our sample.

\subsection{Probabilistic language was discarded in favor of a more deterministic understanding across modalities}
\label{sec:numerical}
Across both LLMs and traditional UIs, clinicians rejected numerical or categorical representation of uncertainty and asked for outputs that pointed to a clear, actionable choice rather than requiring them to interpret or compare probabilities. When the model expressed uncertainty using probabilistic or hedged language, clinicians did not use this uncertainty to guide further reasoning or deliberation. Instead, they desired a single actionable output and abstracted the uncertainty to a single choice based on what they had in mind. Throughout both phases, the model sometimes produced ranked outputs instead of a single clear recommendation. In P3’s session, diagnoses were listed as \textit{``most likely,''} \textit{``less likely,''} and \textit{``rare but possible.''} In P6’s session, treatment options were described using terms such as \textit{``generally preferred,''} \textit{``may be considered,''} and \textit{``associated with a higher likelihood,''} without indicating which option should be chosen. In P10’s session, the model justified its recommendation by referring to \textit{``moderate-quality evidence,''} rather than stating a categorical decision. Across these vignettes, participants did not find this form of uncertainty useful for making a decision. Some clinicians reacted to these forms of uncertainty by trying to collapse them into a yes or no choice. In P3’s session, after scrolling through a ranked list, the participant mentioned, \textit{``Okay, but which one would you actually do?''} and then selected a single option without returning to the model’s probability framing. In P6’s session, the participant said they \textit{``just need to know if this is acceptable or not,''} and moved on without engaging with the rest of the output.

We observed similar behavior during interaction with the UIs. When probability was presented numerically in the UI walkthrough ~\footnote{The model was prompted to express its recommendations using percentage likelihoods and confidence-style values rather than categorical statements.}, participants rejected it and asked for a clearer output. P9 said, \textit{``I don’t know what to do with a 63\% confidence. That doesn’t map to any decision I actually make.''}. P11 added, \textit{``This feels fake-precise. Medicine doesn’t work like that, and I wouldn’t trust myself to act on a number like this''}. P8 similarly stated, \textit{``If it’s not confident enough to tell me yes or no, the number doesn’t help me''}. P10 compared the probability-heavy design to \textit{``a statistics paper, not a clinical tool''} and asked to switch to a version without quantitative confidence annotations.  

\subsection{Keeping all decision factors visible supported comparison, while layering hindered it}
During the UI walkthrough, we examined how different representations and interaction patterns shaped clinicians’ ability to review and compare options under time constraints. We presented 14 candidate variations (see Section~\ref{sec:method}) spanning information structure, quantification, visual encoding, and aggregation, and asked participants to react to each and select favorites.
During the UI walkthrough exercise, participants consistently preferred artifacts that allowed them to see all relevant factors at once and quickly identify a leading option, such as score-based views, cards, and tabular summaries. These artifacts were described as easy to scan and compatible with how clinicians already review charts, notes, and lab results where they have to pay attention to multiple things at once. One clinician said, \textit{``I like that I can see everything at once and immediately know what it’s pushing me toward.''} Another described the format as \textit{``something I could glance at during chart review,''} explaining that \textit{``nothing is hidden, and I don’t have to click around to understand the recommendation.''} A third participant noted that it \textit{“fits how I already scan notes and labs,`` making it “quick to read and easy to compare''}.

\begin{table*}[t]
\centering

\caption{Interaction and Representation Variations Explored in the Study. Participants were asked to pick 3 favorite designs but there was no upper limit to express their positive (+) or negative (-) attitude towards each design. \\ * SHAD: SHape of A Decision}

\label{tab:ui-variations}
\begin{tabular}{lllcc}
\toprule

\textbf{Dimension} & \textbf{Representation Type} & \textbf{UI Variation} & \textbf{Sentiment} \\
\midrule
\multirow{4}{*}{Information Structure}
& Informative & Table & + +   \\
& Informative & Cards & + + +   \\
& Interactive & Zoom &    \\
& Interactive & Path & +   \\
\midrule
\multirow{4}{*}{Quantification}
& Numeric & Scores & + + + + +   \\
& Numeric & Slider &  -  \\
& Non-Numeric & Petals &    \\
& Non-Numeric & Bulb & -   \\
\midrule
\multirow{5}{*}{Visual Encoding}
& Graphical & Bar &    \\
& Graphical & Graph & + + +   \\
& Graphical & Lines &  +  \\
& Abstract & Orbit &    \\
& Abstract & Hexagons  &  \\
\midrule
\multirow{2}{*}{Aggregation}
& Summarized & SHAD* & +   \\
& Structured & Path &    \\
\bottomrule

\end{tabular}
\newline

\end{table*}

Participants were less receptive to artifacts that required interaction to reveal or adjust information. UI designs that relied on zooming, layering, or progressive disclosure were described as disrupting how clinicians review and compare information. One clinician said, \textit{``This might have more detail, but I lose the big picture once I have to start opening things.''} Another remarked, \textit{``If I have to zoom in and out, I’m already annoyed,''} explaining that it made side-by-side comparison harder. A third participant summarized their reaction by saying, \textit{``It’s not that the information is bad. It’s just slower, and I wouldn’t use it in real life.''} At the same time, in about 40\% of encounters clinicians asked at least one follow up after seeing the model’s answer. Interacting through text with the LLM was seen as acceptable because it did not fragment the view of the case or hide information behind controls. Typing or editing a prompt did not interfere with scanning or comparison in the same way that layered or configurable UI elements did.

\subsection{Visual reasoning representations matching with clinicians' thinking style were preferred over text-based descriptions}
Participants consistently preferred visual representations of reasoning over text-based descriptions. Graphs, paths, and decision-tree–like artifacts were described as aligning with familiar clinical tools such as pathways and algorithms. P5 said these views \textit{``feel like how we already think through cases,''} because they could \textit{``see what got ruled out and why something was left standing at the end.''} P7 noted, \textit{``This looks like a clinical pathway. I can trace the reasoning instead of guessing how it got there.''} P10 emphasized that these artifacts supported justification and accountability, stating, \textit{``If I had to explain this decision, I could point to this and walk through the steps.''}. Participants did not use long, narrative explanations of reasoning from the LLM in the same way they used visual reasoning artifacts in the UI. This contrast was explicit in how participants talked about the visual artifacts. P7, after seeing the graph variant, mentioned, \textit{``I can trace the reasoning instead of guessing how it got there,''} indicating that the text outputs left the included reasoning implicit,  rather than inspectable. P5 similarly noted that in the graphs they could \textit{``see what got ruled out and why something was left standing,''} which they did not do when reading narrative model explanations.

\subsection{Voice modality appeared inappropriate to support decision-oriented work}
Voice input was considered unsuitable for creating decision oriented artifacts. While we did not use voice as a modality during the study, clinicians discussed it during interviews in the context of ambient and scribe style systems. P9 described wanting a system that could \textit{``just, like, talk things out loud''} while AI is `\textit{`listening to her and putting all of her thoughts down,'}' and imagined being able to \textit{''take my phone in there… have it listen to my conversation with the patient… and then have it hook up to the computer.''} At the same time, participants raised limits around capturing clinical work through in room recording in front of the patients. P11 noted patient discomfort with being recorded, saying, \textit{``I have had a couple patients be like, you’re recording me, this is weird, I don’t want to do this.''}. The same participant mentioned occasional frustration with the scribing tool as follow:

\begin{quote}

\textit{``I will say it is not… quite so fantastic at, like, capturing a… the physical examination, particularly the musculoskeletal physical examination... it's not catching those things quite as well yet. [...] And then it also has a hard time, like, distinguishing between, like, doctor and patient voice... I'll sometimes ask the patient something or say something, and it kind of interpreted it as, like, the patient was saying it... And then in the history, it'll say, oh, the patient says their hand is swollen... and I'm like, that's not what they said.''}

\end{quote}
Voice was also discussed as a way the AI might return information. P9 described being able to `\textit{`verbally ask a question, and then it’ll give you an answer.''} For decision support, however, spoken output was considered unfit with how clinicians work with different sources of information, such as the notes, the lab results, and the patient at the same time. Unlike on screen outputs that can be skimmed, revisited, or partially ignored, spoken responses force sequential listening and make it harder to jump to a different source of information when it matters. These constraints made voice a poor fit for decision oriented interaction, even when ambient listening was viewed as useful. Participants also pointed to social and practical limits in the exam room, with P11 noting that some patients are \textit{``overly concerned about AI use or, like, surveillance … [and] being surveilled``}.

\section{Design Implications}

The results show that clinicians relied on selective scanning, rejected designs that layered or fragmented information they were comparing, and articulated different expectations for text, visual, and voice-based ambient forms of interaction. Drawing on these observations, we provide three design implications below -  support multitasking and attention management through non-linear information consumption, add interactivity while preserving a holistic, stable, and continuously visible information space, and support coordinated use of different interaction modalities
\subsection{Support multitasking and attention management through non-linear information consumption}
During the study sessions, clinicians worked with AI-generated content by extracting specific cues rather than consuming responses in full. In the text-based LLM sessions, this appeared as skimming for drug names, numbers, or confirming signals and going back to the case or task once those were found. This cue-based scanning and early stopping aligns with bounded rationality in decision making under constraints \cite{Simon1990}, wherein our limited cognitive abilities require us to work with just enough information to make good enough decisions. Consequently, participants reacted most positively to artifacts such as tables, cards, scores, and graph views that made multiple factors visible at once and allowed rapid visual comparison, while designs that implied step by step reading or listening were seen as less compatible with clinical work. These patterns show that clinicians determine when and how any information enters clinical reasoning - in the process of framing their thoughts, they quickly switch attention to finding cues and go back to completing their thought once the cue is found . During the UI walkthroughs, clinicians judged each representation by whether it would let them quickly see what mattered, compare options, and move on. What they valued was not a specific visual style, but whether the artifact supported the same selective, non linear access they used when scanning LLM text, charts, and lab results while reviewing the vignettes. This suggests the need for non-linear layouts unlike voice or text that can enable rapid information scanning and cue spotting.
Decision-making theories applied to clinical judgment from psychology and behavioral science further emphasize this, ranging from normative models based on expected utility and multi-attribute decision theory to descriptive accounts such as fuzzy-trace theory, which emphasizes gist-based reasoning over precise probabilistic calculation \cite{chapman2000decision,reyna2008theories}.

\subsection{Add interactivity while preserving a holistic, stable, and continuously visible information space }
Clinicians did not reject interaction itself. They rejected interaction that fragmented their view of the case and decision-support information. Writing or editing text in the LLM worked because it did not hide information and maintained the information space as is. In contrast, UI designs that required clicking, zooming, or revealing layers were judged by whether they disrupted the ability to keep a consistent information space in view. When interaction meant opening and closing panels, adjusting sliders, or moving between layers, clinicians described losing the big picture and having to mentally reconstruct comparisons that were previously visible at a glance. What mattered was not whether a system was interactive, but what that interaction did to the information space, particularly visibility of the content. Text based prompting let clinicians refine what they asked while still seeing the full case and the full response. Interactive UI controls, by contrast, often traded content visibility for manipulation. Once information had to be uncovered or tuned through controls, clinicians treated it as slower and less useful for decision making, even when the underlying content was the same.
Keeping the full information space visible aligns with clinical cognitive forcing strategies that encourage deliberate checks and guard against premature closure \cite{Croskerry2003}. It also aligns with cognitive forcing functions proposed for human–AI decision support to reduce overreliance on AI \cite{Bucinca2021}.

\subsection{Support coordinated use of different interaction modalities}
Text, visual artifacts, and voice each supported different kinds of work, and none of them worked well on their own. Text conversations with the LLM were useful for asking targeted questions, checking constraints, and retrieving facts, but they did not make the structure of a decision easy to inspect. Visual artifacts such as tables, scores, graphs, and pathways made tradeoffs and eliminations visible, but they could not capture all of the nuance or context that clinicians sometimes needed to express. Voice was attractive for ambient capture and documentation, but it did not fit decision oriented work that required selective attention, comparison, and control over what was revealed.

Clinicians wished to invoke different modes of interaction depending on what they were trying to do. They liked text-based interaction for asking questions or looking up information, preferred visual artifacts when they needed to compare or justify options or understand reasoning behind choices, and identified the need for the system to capture their voice when they wanted it to listen in the background and gather context for better understanding. Clinicians' preferences and behaviors show that these different ways of interacting with AI are not interchangeable ways of doing the same thing. Each modality has a different role for how it is used and affords different ways of entering, examining, and acting on information, and clinicians implicitly treated them as complementary, with different objectives rather than as substitutes.
Treating modalities as complementary implies that each modality supports a distinct decision function, rather than providing multiple interchangeable channels for the same interaction. In this framing, conversational text primarily supports targeted information access and clarification, whereas visual artifacts support deliberation by externalizing the decision space, making alternatives and criteria inspectable, and supporting comparison and revision. This view is consistent with structured human–AI deliberation approaches that keep the current recommendation explicit, make its supporting evidence and criteria visible, and support updating the recommendation when new evidence or constraints are introduced~\cite{Ma2025towards}. When the interface renders alternatives and criteria as explicit objects, it is aligned with decision making procedures studied in multi criteria decision methods such as AHP, where judgments about tradeoffs are expressed as comparisons among options with respect to criteria and can be traced to a resulting prioritization~\cite{saaty2008decision}.

\section{Discussion}

In this paper, we studied how clinicians used an LLM chat interface during complex antibiotic decisions and how they reacted to alternative output formats during a UI walkthrough, alongside interview discussion of voice in ambient scribe systems. We found that clinicians mostly used the model for targeted information lookup and then disengaged without engaging in deliberation or reasoning, preferred data interface formats that supported quick scanning and comparison, and viewed voice as better suited for capture and documentation than for decision support. These observations raise a design question: as models become more capable, what interaction structures help clinicians access that capability without disrupting how decision making is actually carried out in clinical work? While this question has been raised before in the context of traditional machine learning based clinical decision support systems \cite{yang2019CHICDSS}, it needs renewed attention in the context of novel technology like large language models. These models allow for multi-turn interactions as compared to passive consumption of machine recommendations. They also allow clinicians to participate in framing their own decision needs and the corresponding model output as compared to the fixed set of recommendations that the earlier systems offered. We discuss how LLMs with sophisticated natural language and reasoning abilities can move us closer to addressing the above question. 

A key implication is that chat based interaction naturally encourages \textit{``spot checking''} behavior: clinicians can ask a narrow question, scan for the needed cue, and stop, which fits time pressure and lets them keep responsibility for synthesis. That is not a failure mode, it is a rational strategy under uncertainty and workload. The risk is that if the interface mostly presents long, narrative responses, the model’s reasoning remains present but functionally unused, because the output is not shaped for rapid extraction, comparison across options, or easy verification. This creates a future where models improve, but clinical use patterns keep pulling them back toward a faster version of conventional reference tools.

Current research mostly treats LLMs in clinical decision support as turn-taking between a user issuing instructions and a system producing responses, with progress organized as a linear exchange of messages. Recent design guidance for generative AI systems continues to assume this interaction model, emphasizing prompt formulation, iteration, and refinement within a dialogue loop \cite{Weisz2024}. While this paradigm works well for information access and task execution, it makes conversation the default substrate for interaction even in settings where users must integrate information from multiple sources, work with multiple constraints, compare alternatives, and maintain situational awareness. In high-stakes setting of clinical decision-making, this linear, turn-based structure places the burden of coordination on the user, making chat a fragile foundation for joint reasoning rather than a neutral or supportive one.

These observations suggest a gap between what models can do and how clinicians choose to use them in practice. In our study, participants described deliberately keeping the model focused and then extracted only what they needed, which implies that the interaction format shapes expectations about who is responsible for the reasoning and what counts as acceptable support. When outputs are easy to scan, clinicians can stay in control while still having access to deeper content on demand, because they can verify specific points, compare options, and decide what to carry forward into a note, consult, or handoff. The design opportunity is not to make responses longer or more proactive, but to make reasoning easier to request, inspect, and reuse when it is useful, without forcing a fixed, sequential way of consuming information.

Clinicians’ reactions to new AI interaction styles can be understood through the principle of “most advanced, yet acceptable,” a concept pioneered by Raymond Loewy, which shows that people prefer innovations that preserve recognizable features of what they already know \cite{hekkert2003most}. In professional settings, tools do not replace prior practices all at once. They evolve by retaining traces of established artifacts such as notes, tables, and structured pathways while introducing new capabilities. Interfaces that remove these familiar forms in favor of entirely novel interaction styles place users outside their learned routines and create friction, even when the underlying technology is powerful.

\subsection{Limitations and Future Directions}

As with any qualitative study, our findings should be interpreted in light of the boundaries of what we directly observed. First, feedback on the UI variants was elicited through a walkthrough as a set of design probes rather than through hands on use in an interactive system. As a result, these findings reflect clinicians’ interpretation and critique of representations shown during sessions, not behavior that emerges during actual interaction such as navigation, configuration, error recovery, or longer term appropriation. Second, voice was not used as an interaction modality during the tasks. Participants discussed ambient listening and spoken output in interviews, so these results capture perceived workflow and social constraints rather than traces from using voice input or consuming spoken responses during decision work. Additionally, these patterns were observed in a controlled study setting, and their transfer to routine clinical work remains an open question.

This research is a first step toward raising questions about how to move from chat-based information lookup toward interaction structures that better support decision work. One direction is translation from free-text responses into decision representations that can be scanned and compared, for example through structured intermediate ``decision objects'' that can render as tables, cards, or graphs, rather than relying on prompting the model to output numbers. A second direction is to study how models identify and organize the decision itself, including alternatives, constraints, and the specific factors that discriminate between options, so clinicians can request and inspect targeted parts of the reasoning without being pushed into linear reading. A third direction is to develop and evaluate novel UI artifacts for AI-enabled decision making that go beyond current conventional clinical displays, including new visual and interactive primitives for representing trade-offs, uncertainty, and rationale in ways that match clinical attention patterns while still making deeper analysis available on demand.

\subsection{Conclusion}
This qualitative study examines how interaction structures shape clinician–LLM coordination during complex clinical decision making. We used think aloud case walkthroughs and interviews, and we complemented them with a set of UI design probes and follow up reflection to understand how clinicians interpret, verify, and incorporate AI output under realistic constraints. 

Across sessions, clinicians most often treated chat interaction as a tool for targeted retrieval and confirmation, and they frequently disengaged after extracting a single cue. Engagement shifted when the interaction setup positioned the model in a familiar consult role and when reasoning was externalized into stable visual artifacts that supported quick inspection, comparison, and critique, rather than requiring sequential reading. These findings support a view of modalities as complementary, where text supports targeted questioning, visual artifacts support inspecting decision structure and tradeoffs, and voice is better aligned with ambient capture and documentation than decision oriented work. The study also has clear boundaries, since UI feedback was elicited through walkthrough probes rather than hands on deployment, and voice was discussed in interviews rather than used during tasks.


\bibliographystyle{ACM-Reference-Format}
\bibliography{references}

\appendix

\begin{figure*}
    \centering
    \includegraphics[width=0.95\linewidth, alt="A clinical summary for Case A describes a 44-year-old woman with cystic fibrosis and a double lung transplant admitted for sepsis and acute respiratory failure. The vignette details her chief complaint of worsening cough and fever\, physical examination findings including tachycardia and tachypnea\, and investigation results showing acute kidney injury and tacrolimus toxicity. It also lists antimicrobial susceptibilities\, noting that her MRSA infection is susceptible to ceftaroline\, daptomycin\, linezolid\, and vancomycin."]{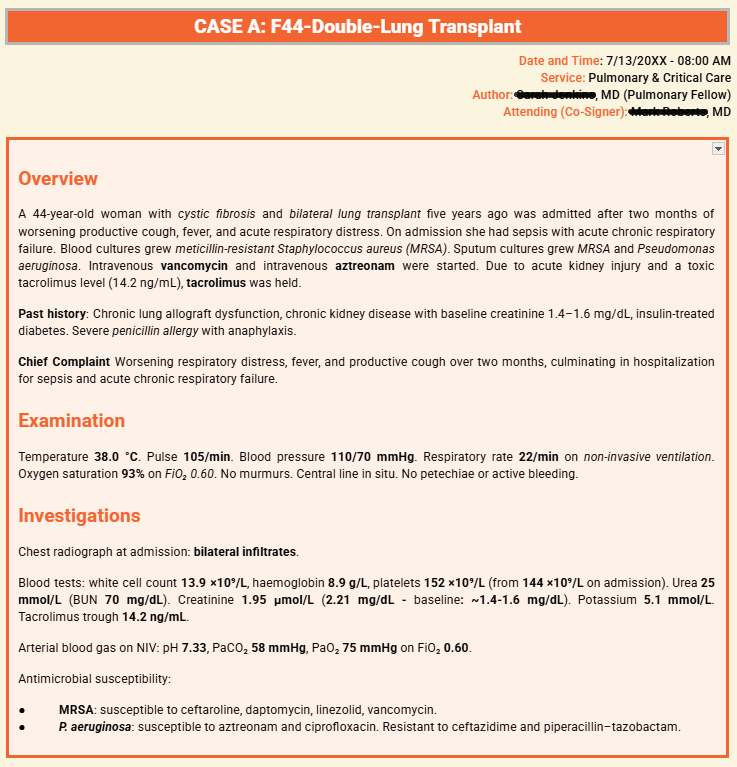}
    \caption{Case A: A clinical summary for Case A describes a 44-year-old woman with cystic fibrosis and a double lung transplant admitted for sepsis and acute respiratory failure.}
    \label{fig:case_a}
\end{figure*}

\begin{figure*}
    \centering
    \includegraphics[width=0.95\linewidth, alt="A clinical summary for Case B describes a 67-year-old man with a prior kidney transplant who presented with septic shock\, acute pulmonary edema\, and acute respiratory failure . The vignette includes a complex medical history of chronic heart failure\, paraplegia\, and stage-4 pressure ulcers with known MRSA and Pseudomonas colonization. Clinical findings show the patient is febrile and tachycardic with an oxygen saturation of 86 percent on room air. Investigations reveal bilateral vascular cephalization on chest X-ray and laboratory results showing an elevated white blood cell count"]{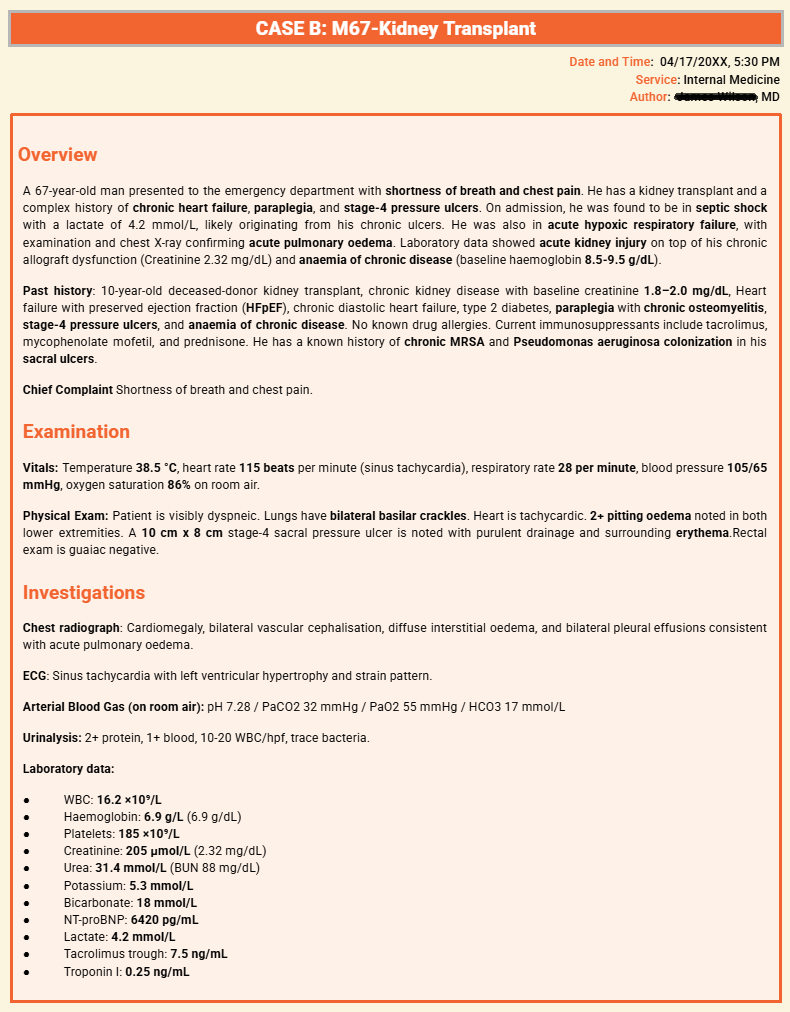}
    \caption{Case B: A clinical summary for Case B describes a 67-year-old man with a prior kidney transplant who presented with septic shock, acute pulmonary edema, and acute respiratory failure .}
    \label{fig:case_b}
\end{figure*}

\begin{table*}
\centering
\caption{Study tasks for Case A and Case B.}
\label{tab:appendix_tasks}
{\ttfamily
\begin{tabular}{p{0.10\linewidth} p{0.12\linewidth} p{0.70\linewidth}}
\hline
\textbf{Case} & \textbf{Task} & \textbf{Task prompt} \\
\hline
Case A & 1 & After initial treatment with intravenous vancomycin, she developed severe nausea and vomiting during admission. What alternative antibiotic are you considering to replace Vancomycin? \\
Case A & 1 & Choose from the following or add your own. Ceftaroline Daptomycin Linezolid Tetracycline Others \\
Case A & 2 & You want to consult your choice of antibiotic in Q1 and the overall antibiotic plan with a specialist. Write a message to the specialist you want to consult. (e.g., infectious disease, transplant pulmonologist, etc.) \\
Case A & 3 & The specialists are not available but you can still ask AI to play the role of specialists and give opinions. In a new AI chat session, please start by saying “You are an expert \_\_\_\_\_\_\_\_\_\_ who has received the following patient case (copy and paste patient case in AI chat).” Now, add your question for the specialist. Then go through the response. \\
Case A & 3-1 & Do you agree or disagree with the AI response? \\
Case A & 3-2 & Do you want to change your previous decision or add new antibiotics? \\
Case A & 4 & For this patient, the treating doctors decided to replace Vancomycin with Linezolid. Why do you think Linezolid was chosen over other antibiotic options? \\
Case B & 1 & Based on the information from the presentation, what is your initial treatment plan? Name the first three actions in their order of priorities. \\
Case B & 2 & The patient's initial treatment in the emergency department consisted of 15 L/min oxygen via non-rebreather mask, one unit of packed red blood cells, and empirical intravenous vancomycin plus piperacillin--tazobactam.. Based on the presentation data, analyze the medical reasoning that justifies this simultaneous, three-part intervention. \\
Case B & 3 & It is the next day and the patient remains febrile (38.2 \textdegree C) and septic, with a high white blood cell count. You have just received the preliminary microbiology results: Blood Culture: Positive for Gram-negative rods. Urine Culture: Positive for >100,000 E. coli. Final Sensitivities: Still pending. The patient is currently still receiving both Vancomycin and Piperacillin-Tazobactam (Zosyn). Given the new culture data and the patient's persistent sepsis, what is your clinical interpretation and what changes will you make to the antibiotic plan? \\
\hline
\end{tabular}
}
\end{table*}

\begin{figure*}
    \centering
    \includegraphics[width=0.95\linewidth, alt="A web-based grid interface titled Transcript Coding Viewer displays coding results with research questions in rows and five AI judges in columns. Each cell contains an AI-generated classification and a link to view specific evidence quotes extracted from the session transcripts."]{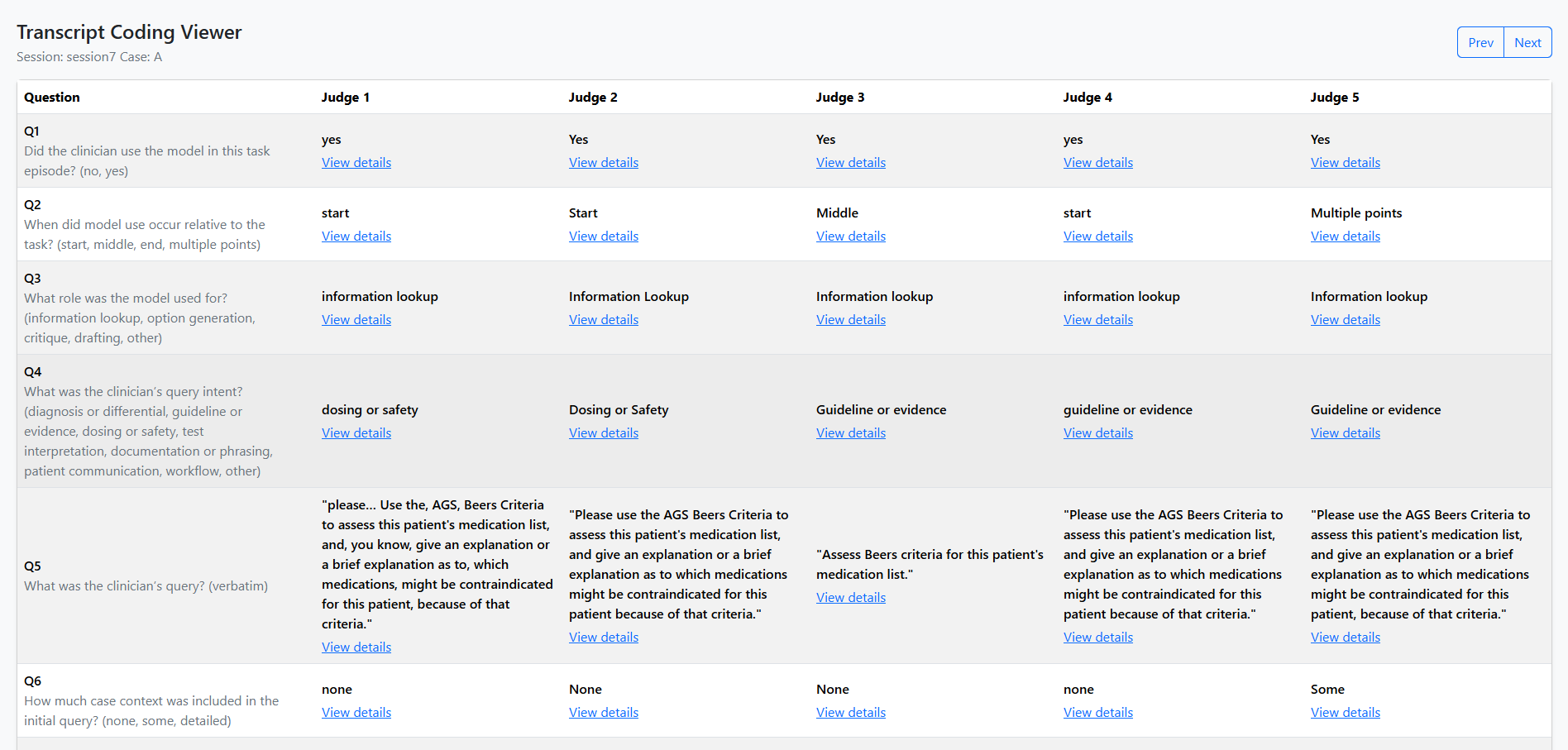}
    \caption{Transcript Coding Viewer interface used by researchers to review and adjudicate parallel AI agent outputs. The grid displays categorical classifications and provides direct links to extracted verbatim evidence for each coding question across five independent AI judges.}
    \label{fig:coding_ui}
\end{figure*}

\begin{table*}
\centering
\caption{Assistant configuration used for AI-assisted coding.}
\label{tab:model_param}
\begin{tabular}{p{0.38\linewidth} p{0.54\linewidth}}
\hline
\textbf{Setting} & \textbf{Value} \\
\hline
Model & gpt-4o \\
Temperature & 1.00 \\
Top P & 1.00 \\
Response format & text \\
Tools: File Search & Enabled (transcript files indexed for retrieval) \\
Tools: Code interpreter & Disabled \\
Functions & None \\
\hline
\end{tabular}
\end{table*}

\begin{figure}
    \centering
    \includegraphics[width=0.95\linewidth, alt="A structured grid titled Agent Comparison Matrix compares different antibiotics in rows against evaluation criteria in columns. The table presents technical clinical data and color-coded status indicators for agents like Linezolid and Ceftaroline to help clinicians compare treatment options at a glance."]{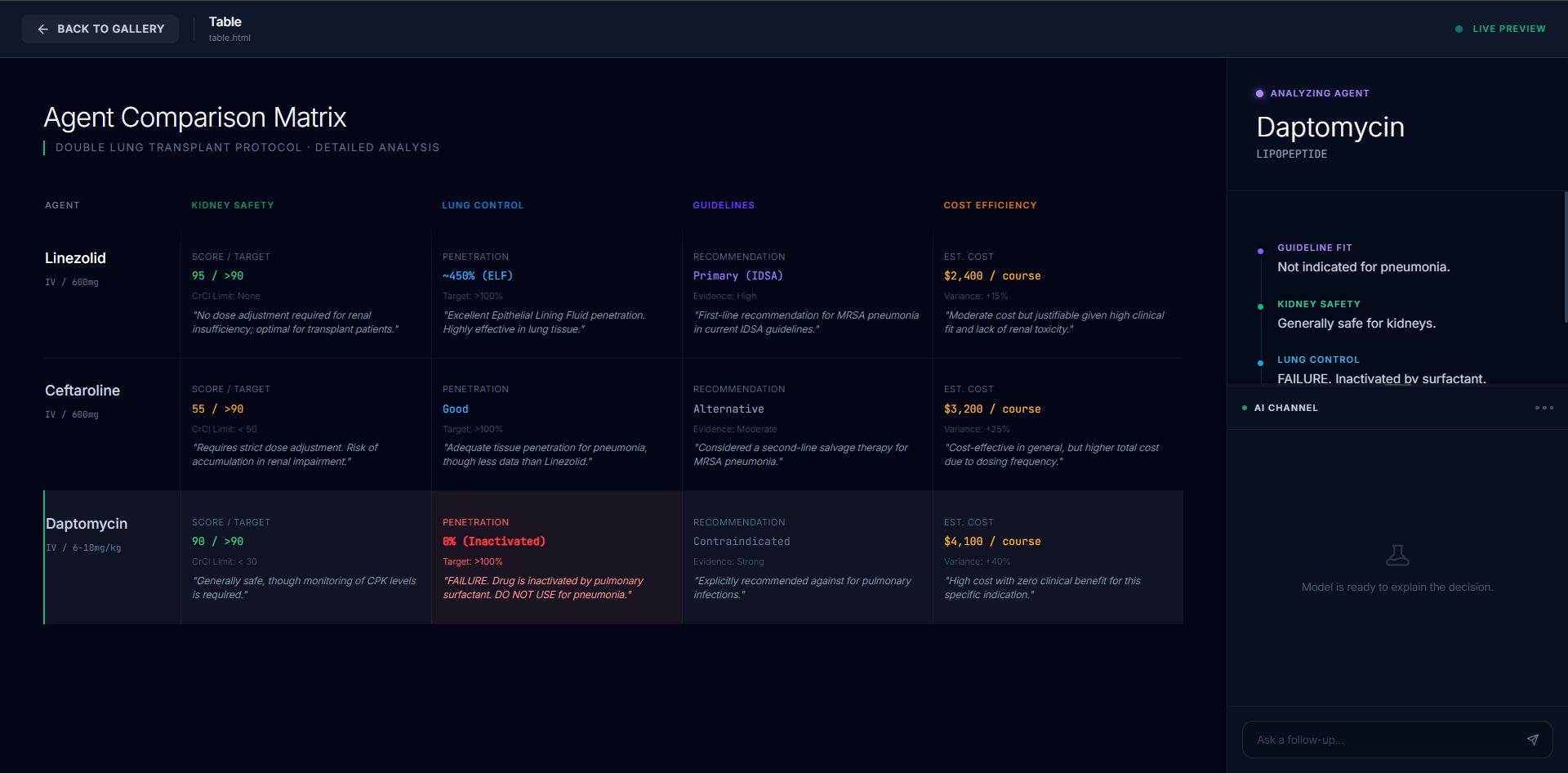}
    \caption{UI variation: \textbf{Table}: This design organizes clinical evaluation criteria such as kidney safety, lung control, and cost efficiency into a traditional comparison matrix to support rapid side-by-side evaluation and identifying a leading clinical option.}
    \label{fig:table}
\end{figure}

\begin{figure}
    \centering
    \includegraphics[width=0.95\linewidth, alt="An interface shows three horizontal stacked bars for the drugs Linezolid\, Ceftaroline\, and Daptomycin. Each bar is divided into colored segments representing different clinical criteria and their relative weight for each antibiotic choice to allow for quick visual comparison."]{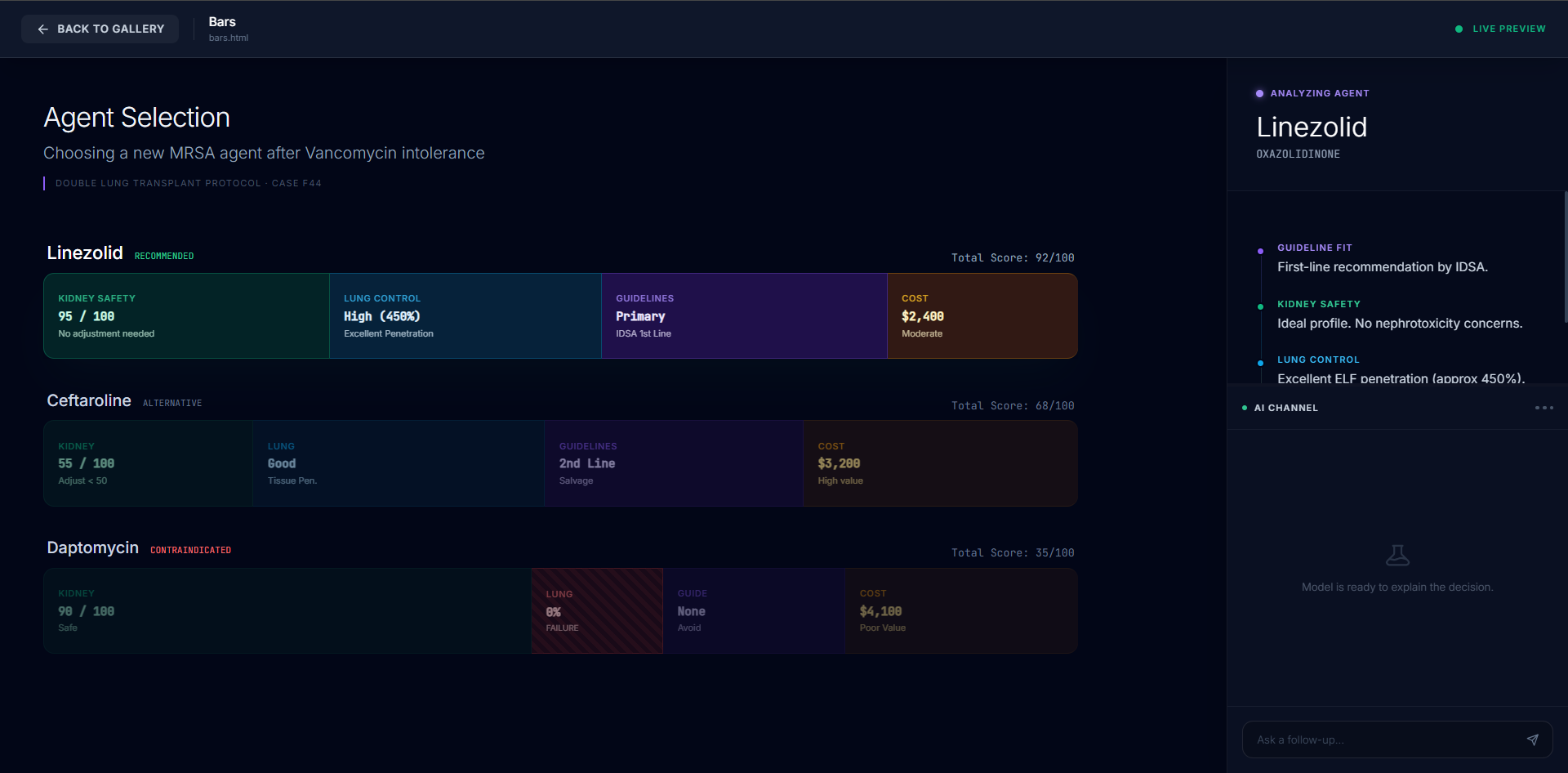}
    \caption{UI variation: \textbf{Bars}. This design organizes clinical decision factors such as kidney safety, lung penetration, and cost into segmented horizontal bars to support rapid comparative reasoning and identifying a leading clinical option.}
    \label{fig:bars}
\end{figure}

\begin{figure}
    \centering
    \includegraphics[width=0.95\linewidth, alt="An interface displaying three circular bubbles for Linezolid\, Ceftaroline\, and Daptomycin. The design requires a user to click or zoom into a specific node to reveal the rationale and clinical data behind each recommendation."]{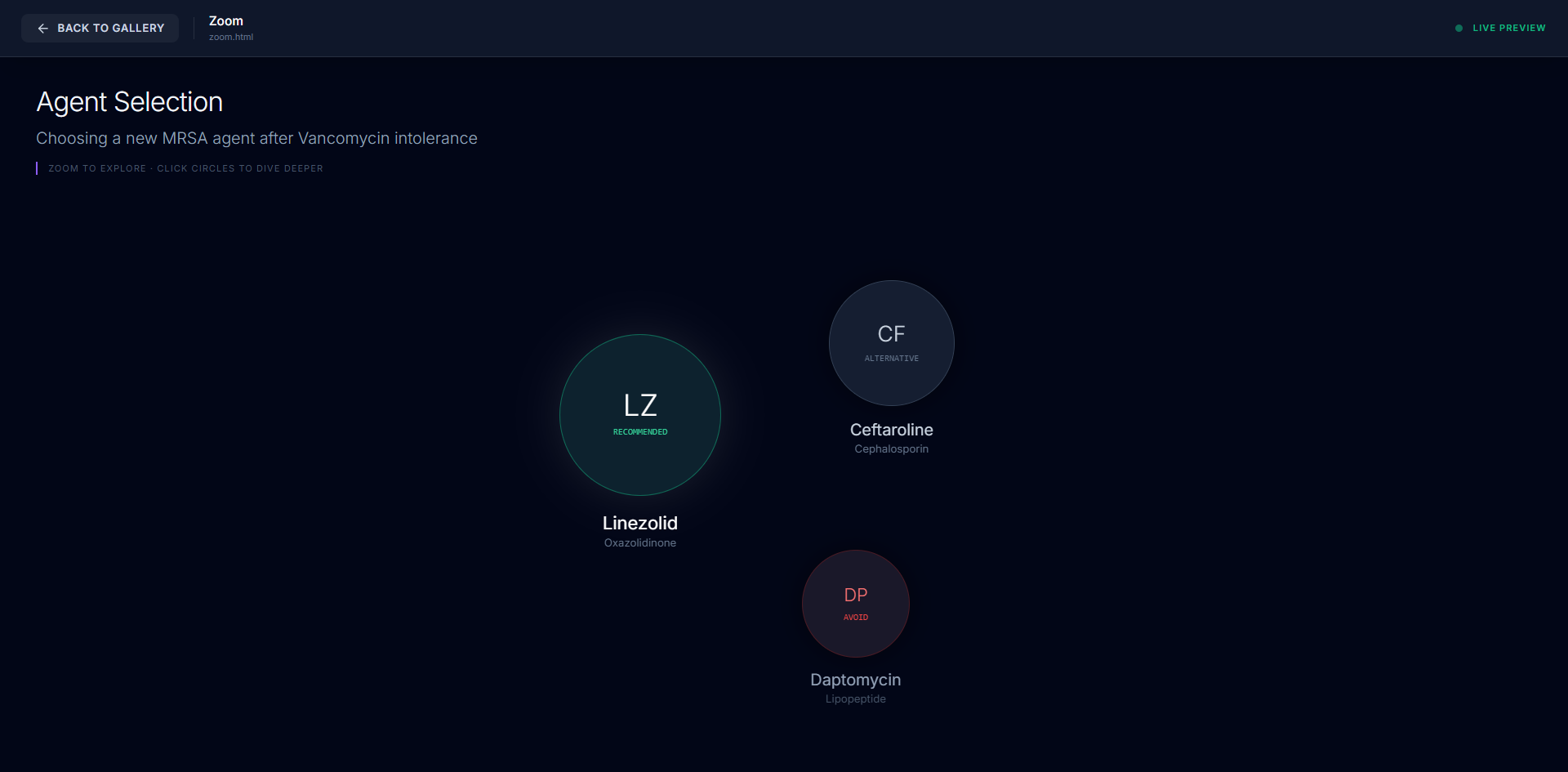}
    \caption{UI variation: Zoom. This design organizes clinical recommendations into interactive circular nodes to support focused exploration of underlying reasoning and evidence.}
    \label{fig:zoom}
\end{figure}

\begin{figure}
    \includegraphics[width=0.95\linewidth, alt="An interface showing a parallel coordinates plot. Antibiotics are represented as individual colored lines that trace across four vertical axes labeled Guidelines\, Kidney Safety\, Lung Control\, and Cost Efficiency. This visualization illustrates the relative performance and trade-offs for each drug option."]{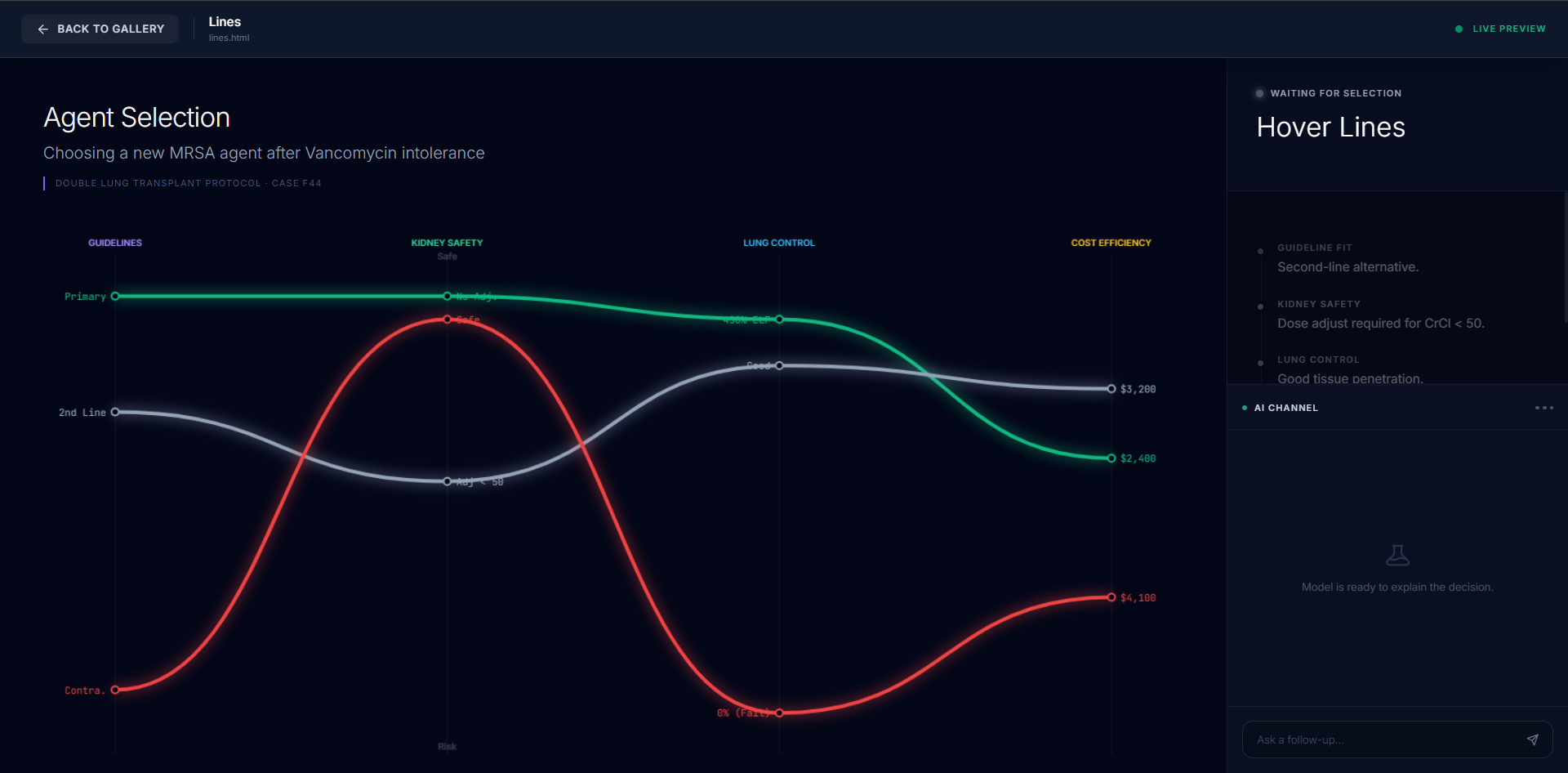}
    \caption{UI variation: \textbf{Lines}. This design organizes clinical decision factors such as guidelines, kidney safety, and lung penetration into vertical parallel axes with colored paths to support rapid comparative reasoning and identifying a leading clinical option.}
    \label{fig:lines}
\end{figure}

\begin{figure}
    \centering 
    \includegraphics[width=0.95\linewidth, alt="An interface titled Analysis: Replacing Vancomycin showing three radar charts for the antibiotics Linezolid\, Daptomycin\, and Ceftaroline. Each drug is represented by a geometric shape plotted across axes for renal safety\, efficacy\, cost\, lung penetration\, and guidelines to visualize the overall profile of each treatment decision."]{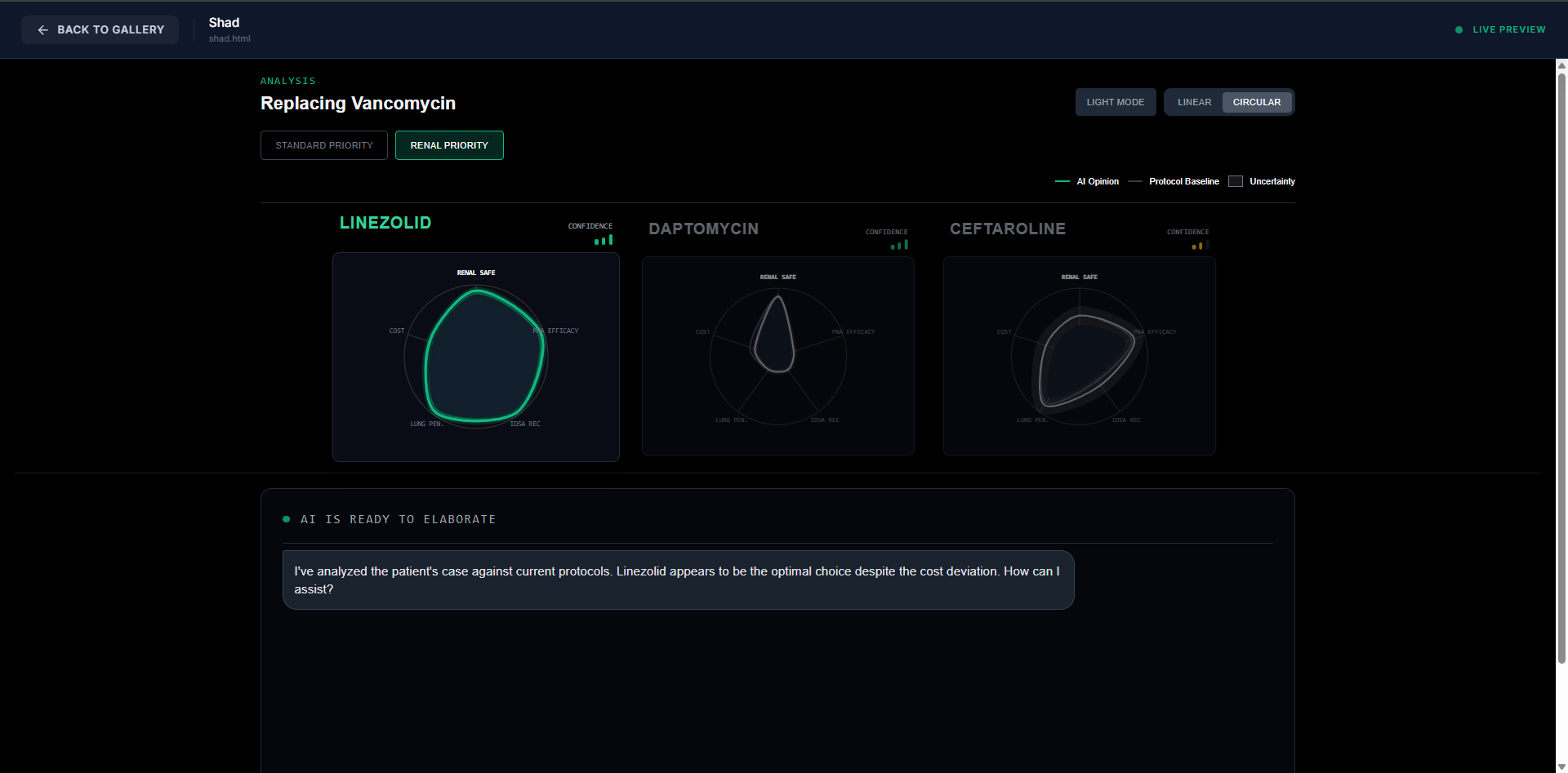}
    \caption{UI variation: \textbf{SHAD}. This design organizes clinical decision factors into radar charts with distinct geometric shapes to support rapid comparative reasoning and identifying a leading clinical option.}
    \label{fig:shad}
\end{figure}

\begin{figure}
    \centering
    \includegraphics[width=0.95\linewidth, alt="An interface showing three colorful glowing shapes for Linezolid\, Ceftaroline\, and Daptomycin. The shapes are positioned relative to clinical dimensions including kidney safety and lung control to visualize drug trade-offs without using numbers."]{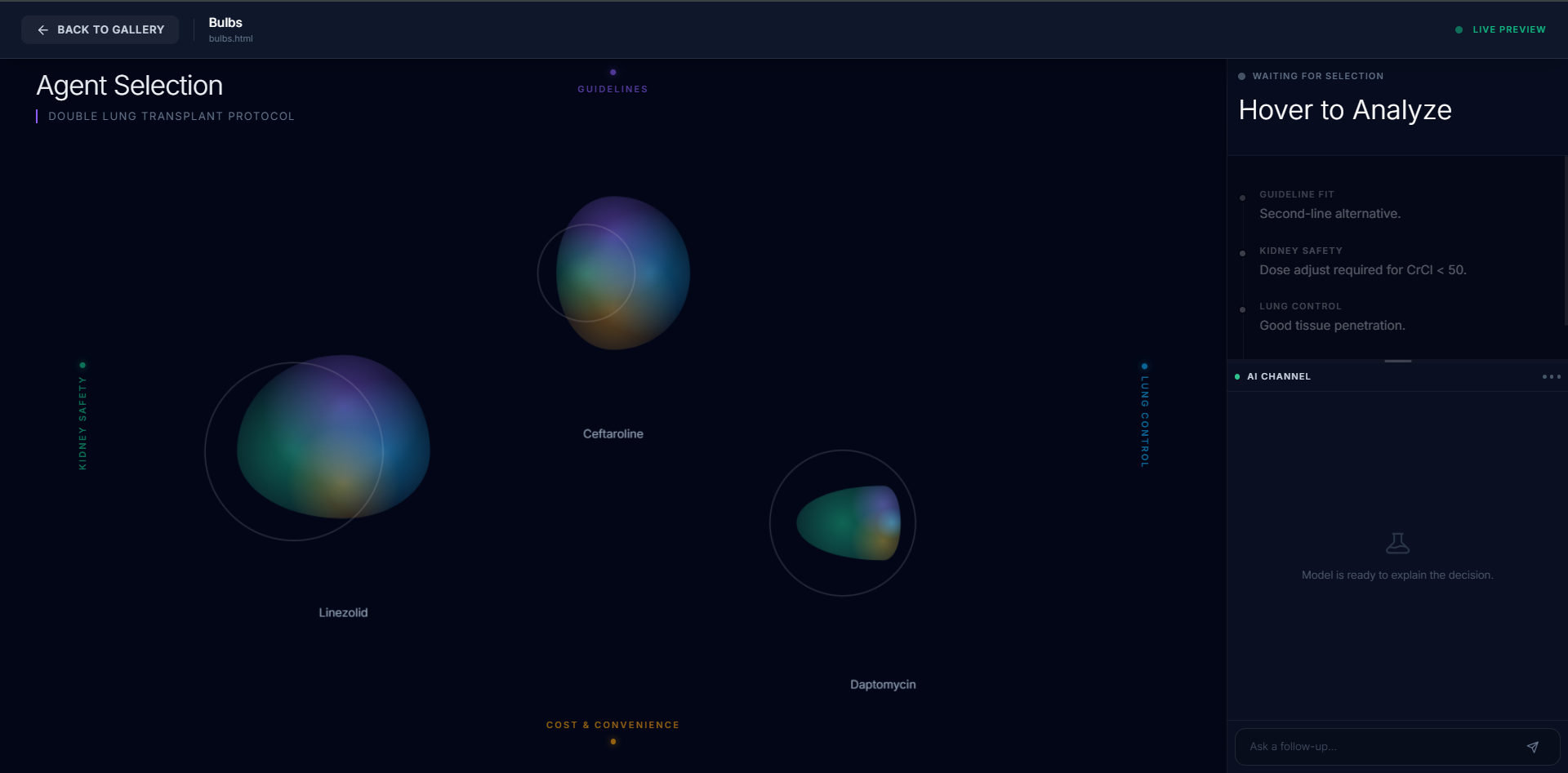}
    \caption{UI variation: \textbf{Bulbs}. This design organizes clinical decision factors into abstract glowing shapes to support qualitative reasoning and identifying a leading clinical option.}
    \label{fig:bulbs}
\end{figure}

\begin{figure}
    \centering
    \includegraphics[width=0.95\linewidth, alt="An interface showing a workspace with large elliptical orbits. Three draggable drug orbs for Linezolid\, Ceftaroline\, and Daptomycin are positioned between axes for Guidelines\, Kidney Safety\, Lung Control\, and Cost. Each orb uses colored rings and beads to visualize clinical data and the importance of specific criteria."]{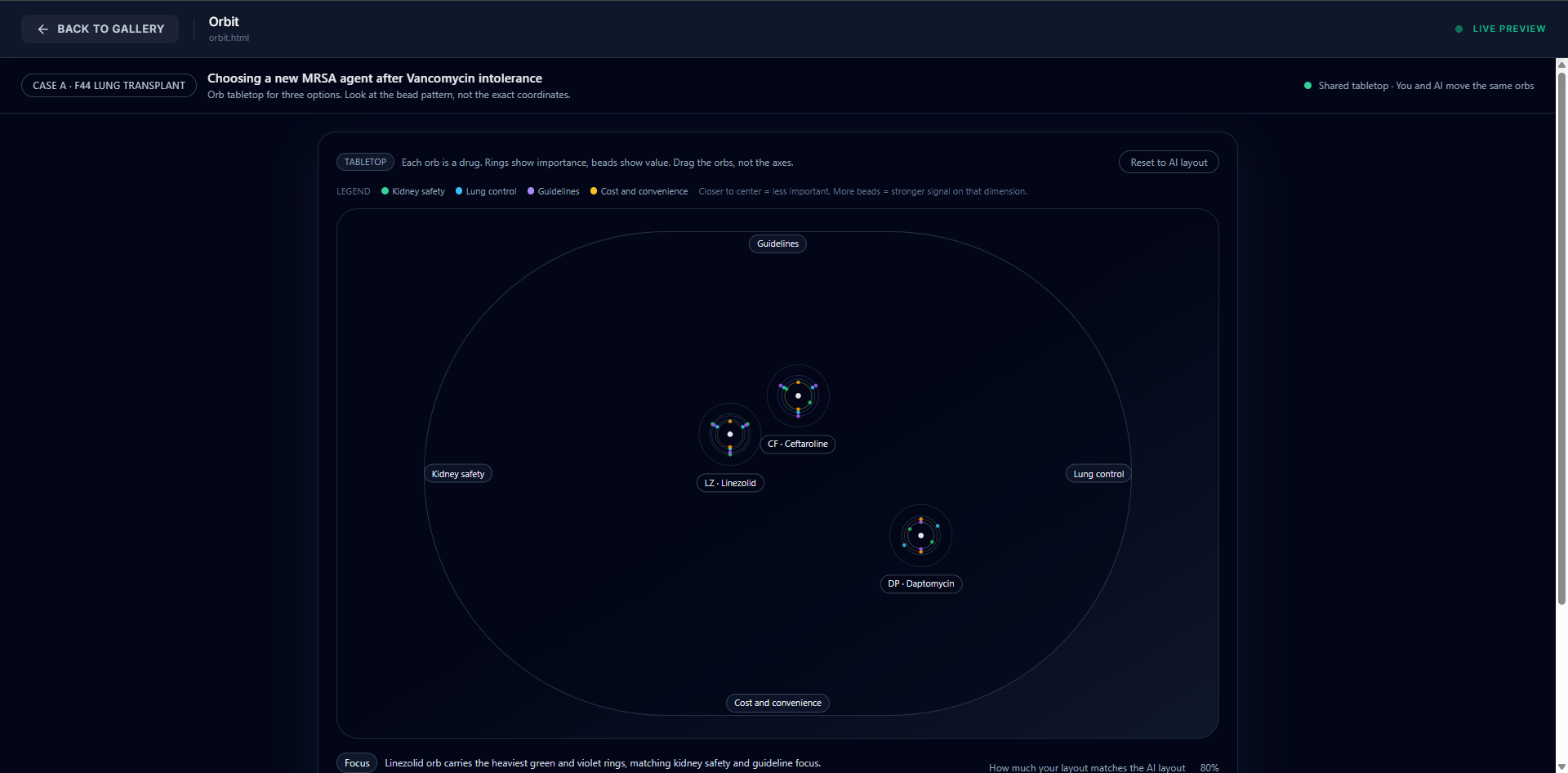}
    \caption{UI variation: \textbf{Orbit}. This design organizes clinical decision factors into an interactive shared tabletop with draggable drug orbs to support rapid comparative reasoning and identifying a leading clinical option.}
    \label{fig:orbit}
\end{figure}

\begin{figure}
    \centering
    \includegraphics[width=0.95\linewidth, alt="Three floral icons for Linezolid\, Ceftaroline\, and Daptomycin. Each drug icon has four colored petals representing different clinical criteria. The size of each petal indicates the strength or importance of that factor for the chosen antibiotic."]{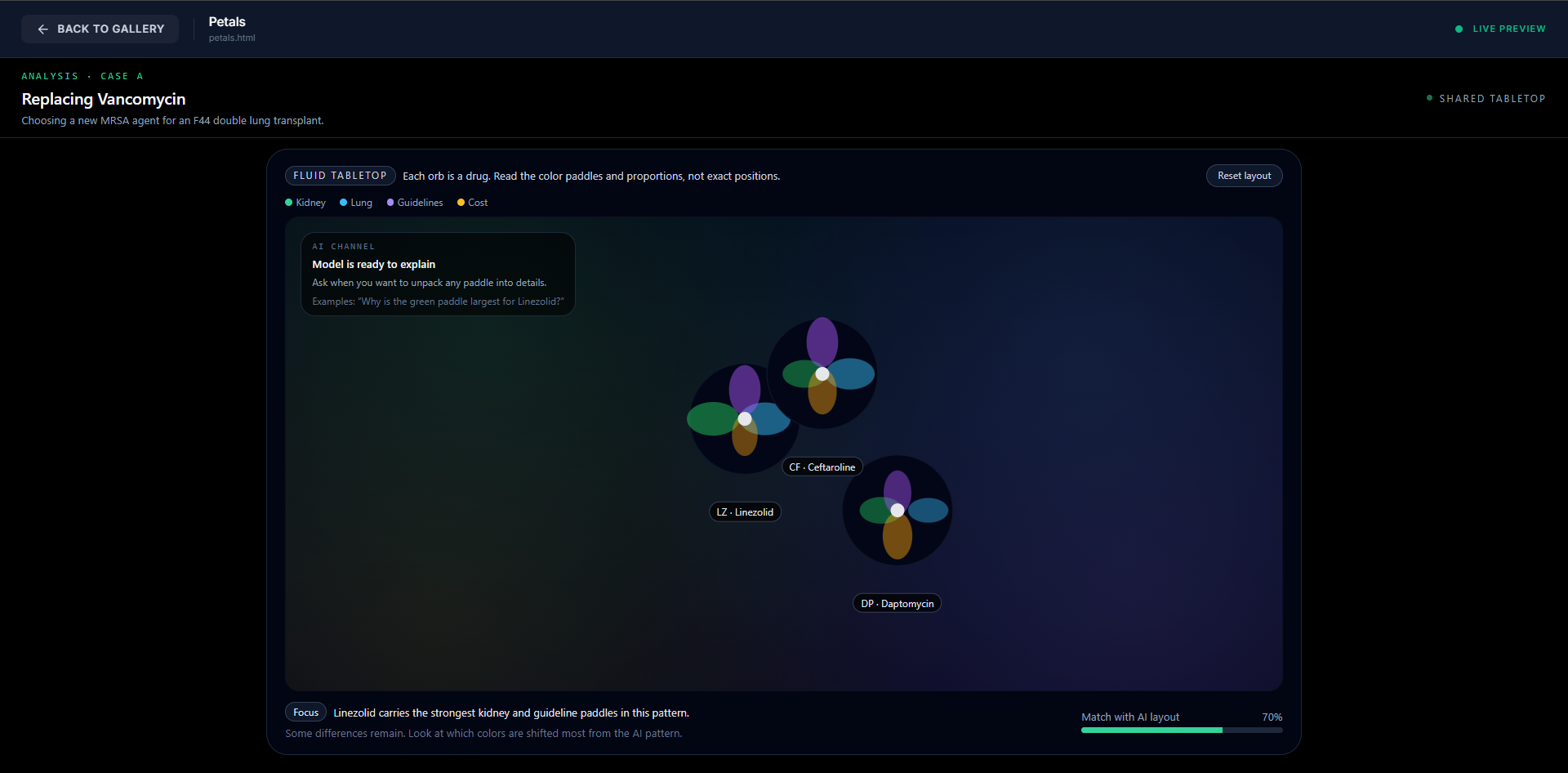}
    \caption{UI variation: \textbf{Petals}. This design organizes clinical decision factors such as kidney safety and guidelines into interactive floral diagrams with colored paddles to support rapid comparative reasoning and identifying a leading clinical option.}
    \label{fig:petals}
\end{figure}

\begin{figure}
    \centering
    \includegraphics[width=0.95\linewidth, alt="An interactive scatter plot with vertical and horizontal axes representing lung coverage and kidney safety. Individual dots representing antibiotics move across the grid as clinical priorities are adjusted. A dashed line connects an AI recommendation to a user selection to highlight the difference between the two treatment choices.
"]{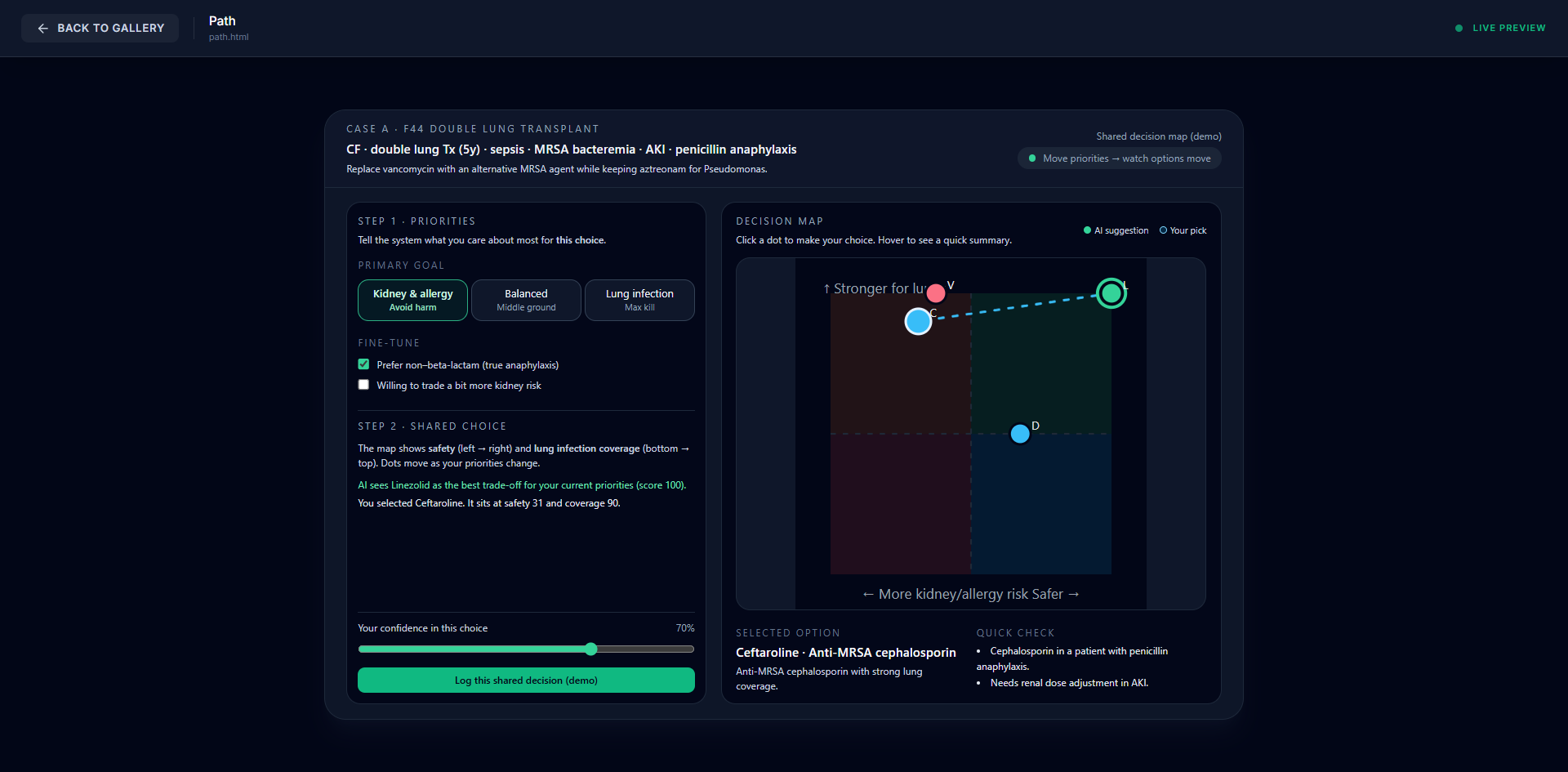}
    \caption{UI variation: \textbf{Path}. This design organizes clinical priorities and drug trade-offs into an interactive scatter plot to support rapid comparative reasoning and identifying a leading clinical option.}
    \label{fig:path}
\end{figure}

\begin{figure}
    \centering
    \includegraphics[width=0.95\linewidth, alt="An interactive medical dashboard with three main sections. The first section features a horizontal slider to adjust the weight between kidney safety and lung infection control. The second section is a scatter plot map showing how drugs like Linezolid and Ceftaroline compare based on those weights. The third section provides a ranked list of treatment options with specific clinical pros and cons for the selected antibiotic.
"]{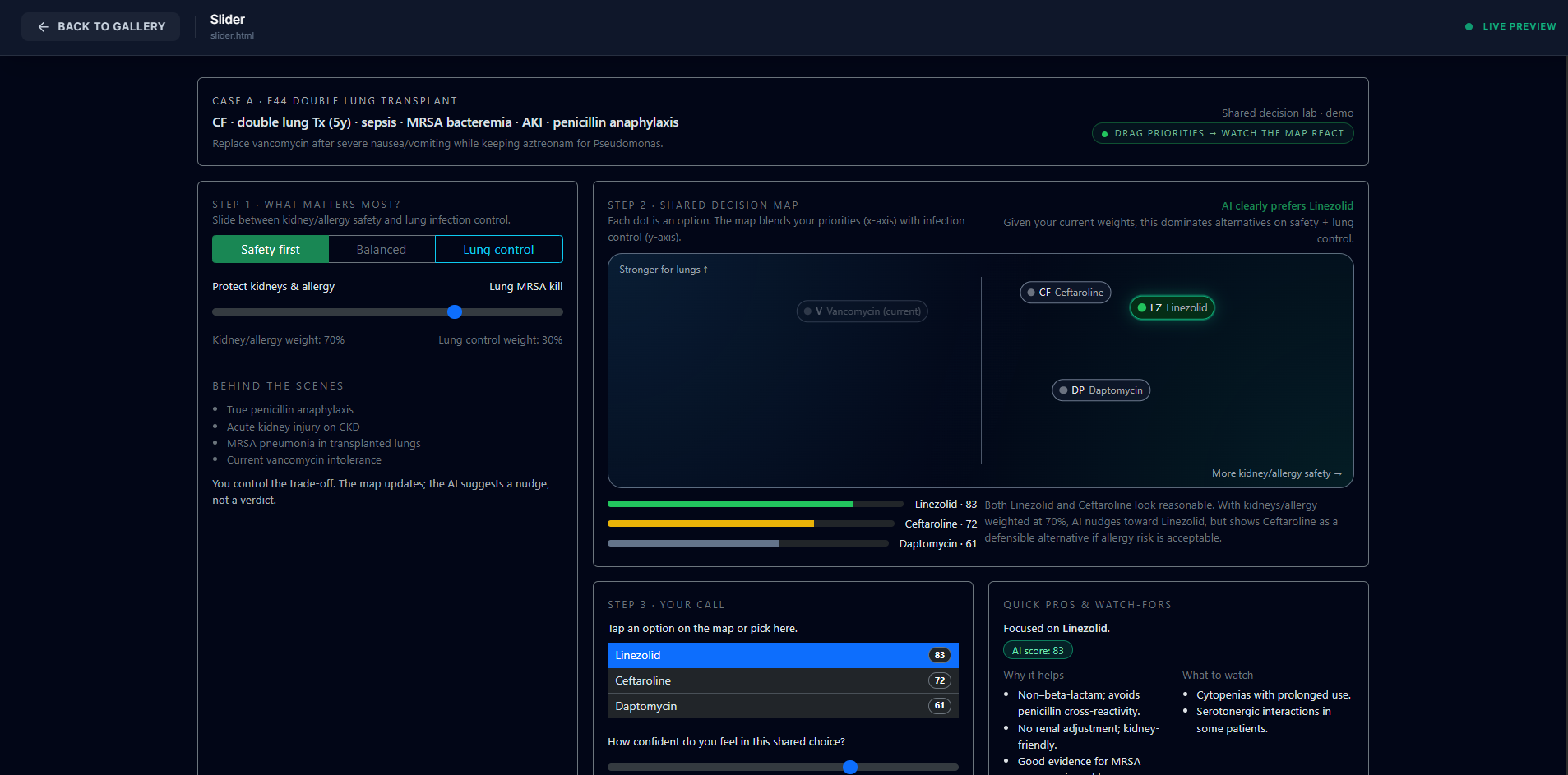}
    \caption{UI variation: \textbf{Slider}. This design organizes clinical priorities and drug trade-offs into an interactive interface with adjustable weight sliders to support rapid comparative reasoning and identifying a leading clinical option.}
    \label{fig:slider}
\end{figure}

\begin{figure}
    \centering
    \includegraphics[width=0.95\linewidth, alt="An interactive interface titled Agent Selection showing three transparent layered planes for the antibiotics Ceftaroline\, Linezolid\, and Daptomycin. Each plane features a central node with colored dots along axes for kidney safety\, lung control\, and guidelines to help clinicians compare individual drug profiles in a 3D layout.
"]{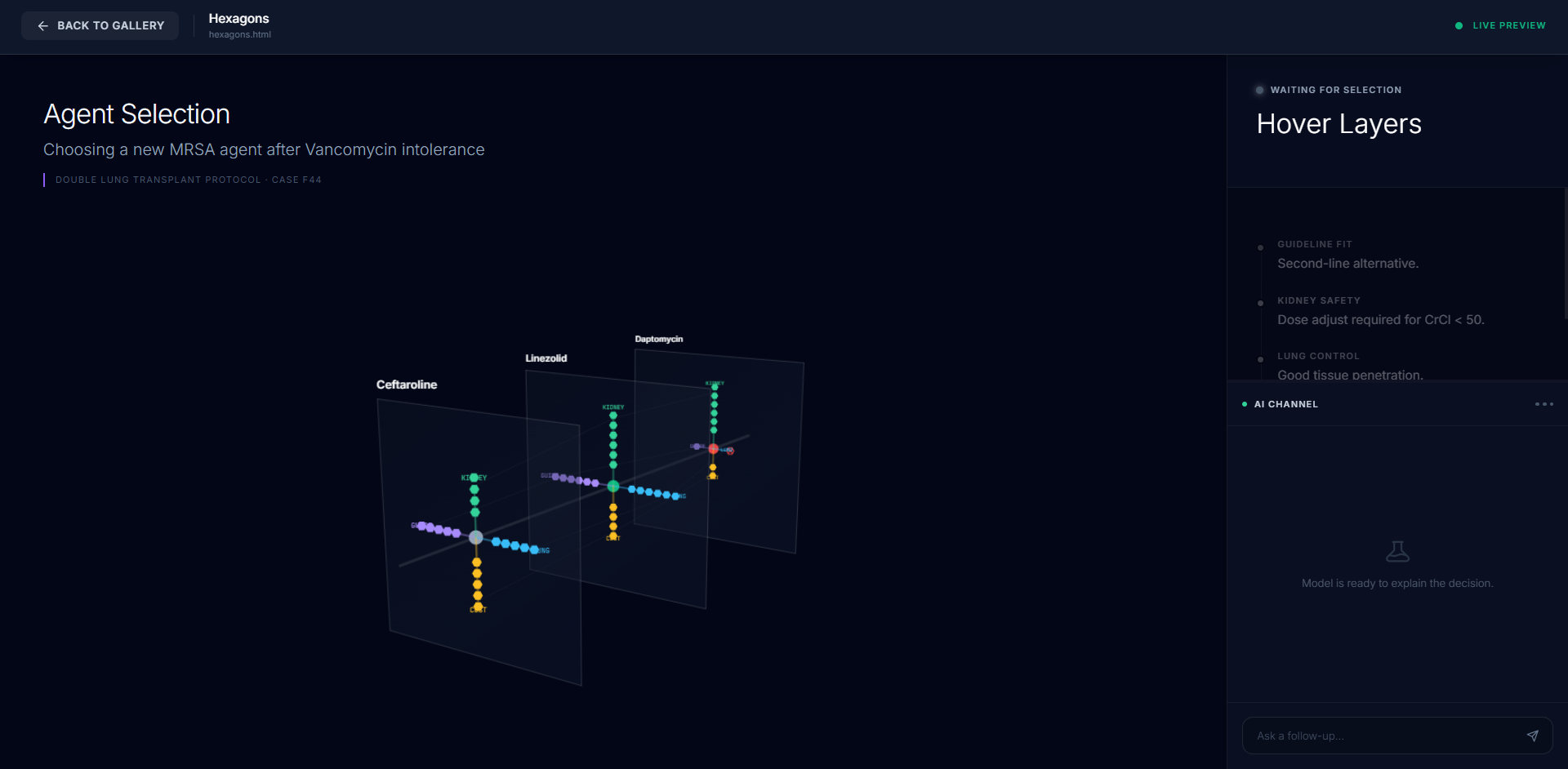}
    \caption{UI variation: Hexagons. This design organizes clinical evaluation criteria into three dimensional layered planes to support rapid comparative reasoning and identifying a leading clinical option.}
    \label{fig:hexagons}
\end{figure}
\end{document}